\documentclass[amsmath,amssymb,prr,aps,showpieces,superscriptaddress,twocolumn,bibnotes,floatfix]{revtex4-2}
\usepackage{amsmath,amsfonts,amssymb,amsthm,graphics,graphicx,epsfig,bbm}
\usepackage[colorlinks=true,citecolor=blue,linkcolor=blue,urlcolor=blue]{hyperref}
\usepackage[usenames]{color}
\usepackage{graphicx}
\usepackage{subfigure}
\usepackage{amsmath}
\usepackage{epsfig}
\usepackage{comment}
\usepackage{dcolumn}
\usepackage{bm}
\usepackage{color}
\usepackage{times}
\usepackage{epstopdf}
\usepackage[dvipsnames]{xcolor}
\usepackage{amssymb}
\usepackage{amstext}
\usepackage{latexsym}
\usepackage{hyperref}
\usepackage{amsfonts}
\usepackage{psfrag}
\usepackage{soul,xcolor}
\usepackage[normalem]{ulem}
\usepackage{dsfont}
\usepackage{txfonts}
\usepackage{float}
\usepackage{booktabs}
\usepackage{dblfloatfix}
\usepackage{caption}
\newcommand{\ket}[1]{\vert #1 \rangle}
\newcommand{\bra}[1]{\langle #1 \vert}

\newcommand{\braket}[2]{\langle #1 \vert #2 \rangle}

\newcommand{\etal}{{\it{et al.}}}

\usepackage{tikz,xcolor,hyperref}
\definecolor{lime}{HTML}{A6CE39}
\DeclareRobustCommand{\orcidicon}{%
	\begin{tikzpicture}
	\draw[lime, fill=lime] (0,0) 
	circle [radius=0.16] 
	node[white] {{\fontfamily{qag}\selectfont \tiny ID}};
	\draw[white, fill=white] (-0.0625,0.095) 
	circle [radius=0.007];
	\end{tikzpicture}
	\hspace{-2mm}
}

\foreach \x in {A, ..., Z}{%
	\expandafter\xdef\csname orcid\x\endcsname{\noexpand\href{https://orcid.org/\csname orcidauthor\x\endcsname}{\noexpand\orcidicon}}
}

\usepackage{orcidlink} 

\renewcommand{\orcidB}[1]{\orcidlink{#1}}

\begin{document}

\setstcolor{red}

\title{Entanglement generation in qubit-ADAPT-VQE through four-qubit algebraic classification}
\date{\today}
\author{D. Tancara \orcidB{0000-0002-5053-3521}} 
\affiliation{Facultad de Física, Pontificia Universidad Católica de Chile, Santiago 7820436, Chile}
\author{H. Díaz-Moraga \orcidB{0009-0007-1446-9060}}
\affiliation{Facultad de Física, Pontificia Universidad Católica de Chile, Santiago 7820436, Chile}
\author{V. Sepúlveda-Trivelli}
\affiliation{Facultad de Física, Pontificia Universidad Católica de Chile, Santiago 7820436, Chile}
\author{D. Goyeneche
\orcidB{0000-0002-9865-4226}}
\affiliation{Facultad de Física, Pontificia Universidad Católica de Chile, Santiago 7820436, Chile}

\begin{abstract}   
While variational quantum algorithms are among the most promising approaches for the noisy intermediate-scale quantum (NISQ) era, their scalability is often hindered by the barren plateau problem. Among the proposals that have demonstrated robustness against this issue, the ADAPT-VQE algorithm stands out for ground state estimation, primarily due to its iterative ansatz construction. Although ADAPT-VQE has been extensively benchmarked on molecular Hamiltonians, where the ground states typically exhibit low entanglement, its performance for highly entangled ground states remains largely unexplored. In this work, we explore a variant of this algorithm known as qubit-ADAPT-VQE, assessing its ability to achieve ground states with substantial entanglement in spin models. We focus on four-qubit systems and employ an algebraic entanglement classification to identify distinct entanglement classes among ground states, and consider a representative of each class as an initial state to evaluate the performance of the algorithm. Our findings highlight the versatility of qubit-ADAPT-VQE, demonstrating that it accurately reaches the ground state across all entanglement classes and initial energy values. 
\end{abstract}
\maketitle

\section{Introduction}

Quantum computing is expected to outperform classical devices in some tasks, as suggested by recent evidence of quantum advantage \cite{AruteNature2019, Zhong2021PhysRevLett, Wu2021PhysRevLett, Madsen2022Nature, Kim2023Nature, Zhu2022SciBull, Morvan2024Nature, Acharya2024Nature, Gao2025PhysRevLett}, even within the limitations of current noisy intermediate-scale quantum (NISQ) devices. This has motivated the development of quantum algorithms specifically designed for the NISQ era, which has become an active area of research \cite{BhartiRevModPhys2022}. This direction is particularly relevant for problems where quantum algorithms are known to provide an advantage, but the required quantum resources are still far beyond what current hardware can support. A clear example is the problem of estimating ground state energies, where quantum phase estimation (QPE) algorithm offers a rigorous approach with potential quantum advantage. However, the large circuit depth required by QPE makes it unsuitable for NISQ devices. For this reason, variational quantum eigensolver (VQE) have emerged as a promising alternative, combining shallow quantum circuits with classical optimization to approximate ground states efficiently in NISQ devices \cite{PeruzzoNatCommun2014}.

In the context of VQE, the choice of ansatz plays a central role, particularly regarding the number of variational parameters and the resulting circuit depth. In a foundational work \cite{PeruzzoNatCommun2014}, an ansatz for molecular simulations was proposed based on the unitary coupled cluster with single and double excitations (UCCSD). This approach involves the exponential of a linear combination of fermionic excitation operators, where the variational parameters correspond to the weights assigned to each excitation term. Another notable approach is the hardware-efficient ansatz \cite{KandalaNature2017}, in which local operations are implemented by using standard quantum gates, and entanglement between qubits is generated through direct hardware-level interactions, following a structure aligned with the underlying device architecture. Across the different ansatz employed in VQE, the barren plateau problem has emerged as a serious challenge and has been extensively investigated in recent years \cite{FontanaNatCommun2024, RagoneNatCommun2024}. This can be characterized by the exponential vanishing variance of the cost function gradients with respect to the system size, which makes variational quantum algorithms increasingly difficult to train as the number of qubits grows. Interestingly, the very constraints that ensure variational quantum algorithms avoid barren plateaus may also suppress their quantum advantage, leading to algorithms that are classically simulable and no longer require a parametrized quantum circuit \cite{CerezoNatCommun2025}. Therefore, the development of algorithms equipped with effective strategies to mitigate barren plateaus remains an active area of research. 

Among the limited set of variational quantum algorithms that are robust to barren plateaus while remaining classically hard to simulate, ADAPT-VQE stands out as a notable example \cite{GrimsleynpjQuantumInf2023, LaroccaNatRevPhys2025}. The main feature of the ADAPT-VQE lies in its dynamic construction of the ansatz \cite{GrimsleyNatCommun2019}. Starting from a reference state, the circuit is built step by step by selecting and appending parameterized unitaries from a predefined operator pool. At each iteration, the selection is guided by the energy gradient associated with each operator, ensuring that the choice reflects the direction of steepest descent in energy, progressively driving the system closer to its ground state.
Initially, ADAPT-VQE was introduced for molecular simulations, where the operator pool is composed of single and double excitations mapped to the qubit representation \cite{GrimsleyNatCommun2019}. These operators, which correspond to those used in the exponential form of the UCCSD ansatz, define the operator pool. By adaptively selecting only the most relevant operators, the algorithm produces significantly shallower circuits than standard UCCSD while achieving higher accuracy, ultimately outperforming this ansatz \cite{GrimsleyNatCommun2019}. Subsequently, an alternative operator pool was introduced, replacing the fermionic single and double excitations mapped to qubit form with direct Pauli strings. This modification defines the qubit-ADAPT-VQE variant \cite{LunTangPRXQuantum2021}, which enables a hardware-efficient ansatz by use of Pauli strings, with improve in circuit depth compared to the original ADAPT-VQE formulation. In the context of molecular simulations, subsequent studies have refined the operator pool used in qubit-ADAPT-VQE by introducing qubit excitations directly, rather than relying on fermionic excitations mapped to qubit representations \cite{YordanovCommunPhys2021}, and by incorporating double qubit excitations within the same spin-orbitals \cite{RamoanpjQuantumInf2025}. 

Outside molecular simulation, ADAPT-VQE has also been applied to problems in another contexts. For instance, an alternative version of the algorithm was employed to prepare the 100-qubit vacuum state of the Schwinger model using a superconducting qubit quantum computer \cite{FarrellPRXQuantum2024}. A modified formulation of ADAPT-VQE was also proposed for quantum dynamics simulation, incorporating McLachlan’s variational principle in the cost function \cite{YaoPRXQuantum2021}. Furthermore, the preparation of Gibbs states using a modified cost function that avoids explicit inclusion of the von Neumann entropy has been addressed through a tailored version of ADAPT-VQE, where a study of entanglement of initial state was also implemented \cite{EconomouIEEE2023}.

Entanglement is a relevant issue in variational quantum algorithms, as it directly influences both the expressiveness of the ansatz and the convergence process. Since the achievable entanglement is fundamentally constrained by the circuit architecture of the ansatz, a mismatch between the required and generated entanglement may lead to inefficient optimization or even convergence failure \cite{WoitzikPhysRevA2020}. While it is often possible to produce large amounts of entanglement within a few circuit layers, excess entanglement, cannot be easily removed and may degrade performance of the algorithm \cite{ChenQuantum2.02023}. In fact, entanglement can lead to barren plateaus  \cite{MarreroPRXQuantum2021}, or can be used for avoiding them \cite{YaoPhysRevA2025} which has motivated extensive investigations of entanglement in various variational quantum algorithms \cite{NakhlPhysRevA2024, WiersemaPRXQuantum2020, DupontPhysRevA2022, SantraPhysRevA2025}.

While entanglement has been investigated in various variational quantum algorithms, its study within the ADAPT-VQE framework remains limited. For instance, entanglement was studied to solve Max-Cut problems in the context of ADAPT-QAOA \cite{ChenQuantum2.02023}, exploring how entanglement arises throughout the procedure, considering that the target solution does not exhibit entanglement. In another work, a modified version of ADAPT-VQE aimed at Gibbs state preparation examined the role of entanglement in the initial state by using graph states \cite{EconomouIEEE2023}, concluding that entangled initial states can provide a better performance.

In this work, we investigate the qubit-ADAPT-VQE algorithm and its capacity to approximate highly entangled ground states in spin systems. Our analysis centers on four-qubit models, where we leverage algebraic entanglement classification together with several entanglement measures to distinguish classes in ground states of models considered. This classification framework allows us to systematically explore how the algorithm performs across different entanglement structures and initial conditions.

\section{Methods}\label{methods}
  In this section we provide the entanglement classification criteria employed for general four-qubit states, as well as the entanglement measures, outline of the quantum spin models considered, and describe the implementation of the qubit ADAPT-VQE algorithm.

\subsection{Algebraic entanglement classification}
In the case of four-qubit systems, entanglement classification has been addressed through a variety of approaches that go beyond the conventional SLOCC framework, which is known to yield an infinite number of inequivalent entanglement classes \cite{DurPhysRevA2000}. An important result was presented in \cite{VerstraetePhysRevA2002}, where it was shown that the infinite number of SLOCC classes can be grouped into nine distinct entanglement families. This foundational work has motivated the development of various alternative entanglement classification schemes for four-qubit systems. Another significant result introduces an algebraic classification of entangled states based on polynomial invariants under the action of the $\text{SL}(2,\mathbb{C})^{\times 4}$ group, the mathematical structure underlying reversible SLOCC transformations \cite{LuquePhysRevA2003}. These invariants are inspired by classical invariant theory, which studies polynomial functions that remain unchanged under group actions. Extending this framework, an alternative approach to entanglement classification has been developed, which is not based on polynomial invariants but instead on the intrinsic algebraic structure of the tensor product space \cite{BuniyJPhysAMathTheor2012}. More precisely, they introduced a collection of discrete invariants determined by the dimensions of the kernels of linear maps between subspaces that compose the entire Hilbert space $\mathcal{H}$. In this work we consider this approach for entanglement classification, proposed in \cite{BuniyJPhysAMathTheor2012_v2,BuniyJPhysAMathTheor2012}.

For four-qubit quantum states are defined on a Hilbert space $\mathcal{H}=\mathcal{H}^{(2)}\otimes\mathcal{H}^{(2)}\otimes\mathcal{H}^{(2)}\otimes\mathcal{H}^{(2)}$, where $\mathcal{H}^{(2)} =\mathbb{C}^2$ denotes the single qubit space. For convenient reasons, the four-qubit space can be seen as a bipartite system  $\mathcal{H}= \mathcal{H}_A\otimes \mathcal{H}_B $, whose elements are given by   
\begin{eqnarray}
\label{eq:bipartite_decomp}
    \ket{\psi} = \sum_{j.k}a_{jk}\ket{j}_A\otimes \ket{k}_B,\label{eq1}
\end{eqnarray}
where $\ket{j}_A\in \mathcal{H}_A $ and $\ket{k}_B\in \mathcal{H}_B$. Here, we can introduce linear maps of the form $f_{\ket{\psi}}: \mathcal{H}_A \rightarrow \mathcal{H}_B$, acting on a state $\ket{\omega}_A \in \mathcal{H}_A$, that takes the form
\begin{eqnarray}
    f_{\ket{\psi}}(\ket{\omega}_A) &=& (\bra{\omega}_A\otimes \mathbb{I}_B)\ket{\psi}\nonumber\\
    &=& \sum_{j,k}a_{jk}\braket{\omega}{j}_A \ket{k}_B\nonumber\\
    &=& \sum_{k}b_{k}\ket{k}_B,
\end{eqnarray}
where $b_{k} = \sum_j a_{jk}\braket{\omega}{j}_A$. Both kernel and image of a linear map are two fundamental subspaces that capture essential structural properties of the map. Nonetheless, one can focus on the kernel, as the image is the unique orthogonal complement of the kernel, according to linear algebra. Considering the kernel as
\begin{eqnarray}
    \text{ker}(f_{\ket{\psi}}) = \{\ket{\omega}_A \in \mathcal{H}_A:  f_{\ket{\psi}}(\ket{\omega}_A) = 0  \}, 
\end{eqnarray}
note that the dimension of the kernel is invariant under the action of the group $\text{GL}(2,\mathbb{C})^{\times 4}$. By this reason it can be used to define entanglement invariants. The complete set of invariants corresponds to the sequence of dimensions of all possible intersections of the elements of $\ker{(f_{\ket{\psi}}\otimes\mathbb{I}_B})$, as this makes the classification unique \cite{BuniyJPhysAMathTheor2012}. This methods provides 19 independent invariants that characterizes four qubit states into 83 different classes, which reduce to 27 up to permutations of qubit. Table \ref{table1} shows the 27 entanglement classes and a representative state for each of them, following the labeling convention used in \cite{BuniyJPhysAMathTheor2012}. More details about algebraic entanglement classes can be found in Appendix \ref{appendix1}.
\begin{widetext} 
\centering
\begin{tabular}{|c|c|}
\hline
\textbf{Class} & \textbf{Representative state 
$\ket{C_l}$} \\
\hline
$C_0$ & 0 \\
\hline
$C_1$  &   $\ket{0000}$ \\
\hline
$C_3$  &   $\frac{1}{\sqrt{2}} (\ket{0000} + \ket{1010})$ \\
\hline
$C_9$  &   $\frac{1}{2} (\ket{0000} + \ket{0101} + \ket{1010} + \ket{1111})$ \\
\hline
$C_{11}$ &   $\frac{1}{\sqrt{3}} (\ket{0000} + \ket{1010} + \ket{1100})$ \\
\hline
$C_{15}$ &   $\frac{1}{\sqrt{2}} (\ket{0000} + \ket{1110})$ \\
\hline
$C_{19}$ &   $\frac{1}{2} (\ket{0000} + \ket{1001} + \ket{1010} + \ket{1100})$ \\
\hline
$C_{22}$ &   $\frac{1}{\sqrt{3}} (\ket{0000} + \ket{0101} + \ket{1110})$ \\
\hline
$C_{26}$ &   $\frac{1}{\sqrt{2}} (\ket{0000} + \ket{1111})$ \\
\hline
$C_{28}$ &   $\frac{1}{\sqrt{5}} (\ket{0000} + \ket{0011} + \ket{0110} + \ket{1001} + \ket{1100})$ \\
\hline
$C_{31}$ &   $\frac{1}{\sqrt{12}} (\ket{0000} - \ket{0001} - \ket{0010} - \ket{0100} + \ket{0101} + \ket{0111} - \ket{1000} + \ket{1010} + \ket{1011} + \ket{1101} + \ket{1110} + \ket{1111})$ \\
\hline
$C_{33}$ &   $\frac{1}{\sqrt{6}} (\ket{0000} + c\ket{0011} -(1+c)\ket{0101} -(1+c)\ket{1010} + c\ket{1100} + \ket{1111})$, $c\in \mathbb{R} \setminus \{-2,-1,0\}$ \\
\hline
$C_{36}$ &   $\frac{1}{\sqrt{4}} (\ket{0000} + \ket{0010} + \ket{0111} + \ket{1001})$ \\
\hline
$C_{39}$ &  $\frac{1}{\sqrt{3}} (\ket{0000} + \ket{0101} + \ket{1010})$ \\
\hline
$C_{43}$ &   $\frac{1}{\sqrt{5}} (\ket{0000} + \ket{0001} + \ket{0100} + \ket{1011} + \ket{1110})$ \\
\hline
$C_{48}$ &   $\frac{1}{\sqrt{4}} (\ket{0000} + \ket{0111} + \ket{1010} + \ket{1101})$ \\
\hline
$C_{50}$ &   $\frac{1}{\sqrt{4}} (\ket{0000} + \ket{0111} + \ket{1010} + \ket{1100})$ \\
\hline
$C_{57}$ &   $\frac{1}{\sqrt{4}} (\ket{0000} + \ket{0011} + \ket{0110} + \ket{1100})$ \\
\hline
$C_{60}$ &   $\frac{1}{\sqrt{6}} (\ket{0000} + \ket{0101} + \ket{0110} + \ket{1001} + \ket{1010} + \ket{1111})$ \\
\hline
$C_{62}$ &   $\frac{1}{\sqrt{6}} (\ket{0000} + \ket{0001} + \ket{0110} + \ket{1000} + \ket{1011} + \ket{1100})$ \\
\hline
$C_{65}$ &   $\frac{1}{\sqrt{10}} (\ket{0000} + \ket{0001} + \ket{0011} + \ket{0101} + \ket{0110} + \ket{1000} + \ket{1010} + \ket{1101} + \ket{1110} + \ket{1111})$ \\
\hline
$C_{67}$ &   $\frac{1}{\sqrt{9}} (\ket{0000} + \ket{0001} + \ket{0010} - \ket{0100} + \ket{0101} - \ket{1000} + \ket{1010} + \ket{1100} + \ket{1111})$ \\
\hline
$C_{68}$ &   $\frac{1}{\sqrt{5}} (\ket{0000} + \ket{0010} + \ket{0100} + \ket{1000} + \ket{1111})$ \\
\hline
$C_{72}$ &   $\frac{1}{\sqrt{5}} (\ket{0000} + \ket{0111} + \ket{1001} + \ket{1010} + \ket{1100})$ \\
\hline
$C_{75}$ &   $\frac{1}{\sqrt{5}} (\ket{0000} + \ket{0011} + \ket{0100} + \ket{1001} + \ket{1110})$ \\
\hline
$C_{80}$ &   $\frac{1}{\sqrt{6}} (\ket{0000} + \ket{0010} + \ket{0101} + \ket{1000} + \ket{1011} + \ket{1110})$ \\
\hline
$C_{82}$ &   $\frac{1}{\sqrt{7}} (\ket{0000} + \ket{0011} + \ket{0100} + \ket{0101} + \ket{1000} + \ket{1010} + \ket{1111})$ \\
\hline
\end{tabular}
\captionof{table}{Entanglement classes and representative states for four-qubit states \cite{BuniyJPhysAMathTheor2012}.}\label{table1}
\end{widetext}
\subsection{Entanglement measures}
Following the algebraic classification of entanglement into 27 distinct classes, subsequent research introduced a complementary framework based on continuous entanglement invariants for the same classes \cite{JafarizadehPhysLettA2019}. These invariants corresponds to entanglement measures and classical invariants, forming a set of 11 continuous quantities. Let $\mathcal{H}_l = \mathbb{C}^2, l =\{A, B, C, D\}$ be the Hilbert Space associated to the first, second, third and fourth qubit, respectively, the continuous invariants considered are:
\begin{itemize}
    \item Concurrence between two qubits in the state $\rho_{AB} = \text{Tr}_{BC}(\ket{\psi}\bra{\psi})$:
    \begin{eqnarray}
        C_{AB} = \text{max}_{\rho_{AB}}\{0, \lambda_1-\lambda_2-\lambda_3-\lambda_4\},
    \end{eqnarray}
    where $\lambda_i$ are eigenvalues in descendent order of the operator:
    \begin{eqnarray}
        R = \sqrt{\sqrt{\rho_{AB}}\tilde{\rho}_{AB}\sqrt{\rho_{AB}}},
    \end{eqnarray}
    and $\tilde{\rho}_{AB}$ is the complex conjugate of $\rho_{AB}$ in the computational basis. Considering all pair of qubits we have the further invariants $C_{AC}, C_{AD}, C_{BC}, C_{BD}$ and $C_{CD}$.

    \item Genuine multipartite (GM) entanglement is quantified by    \begin{eqnarray}
        C_{GM}(\ket{\psi}) = \text{min}_{P\in (A|B)} \sqrt{2(1-\rho_A^2)},
    \end{eqnarray}
    where $(A|B)$ are all possible bipartitions of four-qubit system and $\rho_A = \text{Tr}_B(\ket{\psi}\bra{\psi})$. This  measure detects whether each qubit is entangled with all the rest.
    \item Polynomial invariants for a general four-qubit state $\ket{\psi} = \sum_{jklm}v_{j,k,l,m}\ket{jklm}$, defined  in the computational basis \cite{LuquePhysRevA2003}:
    \begin{eqnarray}
h_1 &=&
v_{0,0,0,0}v_{1,1,1,1}
- v_{0,0,0,1}v_{1,1,1,0}
- v_{0,0,1,0}v_{1,1,0,1} \\ \notag
&&
+ v_{0,0,1,1}v_{1,1,0,0} - v_{0,1,0,0}v_{1,0,1,1}
+ v_{0,1,0,1}v_{1,0,1,0} \\ \notag
&& + v_{0,1,1,0}v_{1,0,0,1}
- v_{0,1,1,1}v_{1,0,0,0}
\end{eqnarray}

\begin{eqnarray}
h_2 &=& \left|
\begin{array}{cccc}
v_{0,0,0,0} & v_{0,1,0,0} & v_{1,0,0,0} & v_{1,1,0,0} \\
v_{0,0,0,1} & v_{0,1,0,1} & v_{1,0,0,1} & v_{1,1,0,1} \\
v_{0,0,1,0} & v_{0,1,1,0} & v_{1,0,1,0} & v_{1,1,1,0} \\
v_{0,0,1,1} & v_{0,1,1,1} & v_{1,0,1,1} & v_{1,1,1,1}
\end{array}
\right|
\end{eqnarray}

\begin{eqnarray}
h_3 &=& \left|
\begin{array}{cccc}
v_{0,0,0,0} & v_{1,0,0,0} & v_{0,0,1,0} & v_{1,0,1,0} \\
v_{0,0,0,1} & v_{1,0,0,1} & v_{0,0,1,1} & v_{1,0,1,1} \\
v_{0,1,0,0} & v_{1,1,0,0} & v_{0,1,1,0} & v_{1,1,1,0} \\
v_{0,1,0,1} & v_{1,1,0,1} & v_{0,1,1,1} & v_{1,1,1,1}
\end{array}
\right|
\end{eqnarray}
\begin{eqnarray}
h_4 &=& \det(h_{4,jk})_{1 \leq j,k \leq 3}, \quad \text{where:} \nonumber \\
h_{4,1,1} &=& -v_{0,0,0,1}v_{0,0,1,0} + v_{0,0,0,0}v_{0,0,1,1} \nonumber \\
h_{4,1,2} &=& v_{0,0,1,1}v_{0,1,0,0} - v_{0,0,1,0}v_{0,1,0,1}
- v_{0,0,0,1}v_{0,1,1,0} \nonumber \\
&& + v_{0,0,0,0}v_{0,1,1,1} \nonumber \\
h_{4,1,3} &=& -v_{0,1,0,1}v_{0,1,1,0} + v_{0,1,0,0}v_{0,1,1,1} \nonumber \\
h_{4,2,1} &=& v_{0,0,1,1}v_{1,0,0,0} - v_{0,0,1,0}v_{1,0,0,1}
- v_{0,0,0,1}v_{1,0,1,0} \nonumber \\ 
&& + v_{0,0,0,0}v_{1,0,1,1} \nonumber \\
h_{4,2,2} &=& v_{0,1,1,1}v_{1,0,0,0} - v_{0,1,1,0}v_{1,0,0,1}
- v_{0,1,0,1}v_{1,0,1,0} \nonumber \\ 
&& + v_{0,1,0,0}v_{1,0,1,1} + v_{0,0,1,1}v_{1,1,0,0} - v_{0,0,1,0}v_{1,1,0,1} \nonumber \\
&& - v_{0,0,0,1}v_{1,1,1,0} + v_{0,0,0,0}v_{1,1,1,1} \nonumber \\
h_{4,2,3} &=& v_{0,1,1,1}v_{1,1,0,0} - v_{0,1,1,0}v_{1,1,0,1}
- v_{0,1,0,1}v_{1,1,1,0} \nonumber \\ 
&& + v_{0,1,0,0}v_{1,1,1,1} \nonumber \\
h_{4,3,1} &=& -v_{1,0,0,1}v_{1,0,1,0} + v_{1,0,0,0}v_{1,0,1,1} \nonumber \\
h_{4,3,2} &=& v_{1,0,1,1}v_{1,1,0,0} - v_{1,0,1,0}v_{1,1,0,1}
- v_{1,0,0,1}v_{1,1,1,0} \nonumber \\ 
&& + v_{1,0,0,0}v_{1,1,1,1} \nonumber \\
h_{4,3,3} &=& -v_{1,1,0,1}v_{1,1,1,0} + v_{1,1,0,0}v_{1,1,1,1}.
\end{eqnarray}    
\end{itemize}
Therefore, the set of continuous invariants
\begin{eqnarray*}  
(h_1,h_2,h_3,h_4, C_{AB}, C_{AC}, C_{AD}, C_{BC}, C_{BD}, C_{CD}, C_{GM})
\end{eqnarray*}
allows us to classify the 27 entanglement classes \cite{BuniyJPhysAMathTheor2012}, each of them uniquely identified by a specific combination of non-vanishing invariants, as shown in Table \ref{table2}.
\begin{table}[H]
\centering
\begin{tabular}{lcccccccccccc}
\toprule
Class & $h_1$ & $h_2$ & $h_3$ & $h_4$ & $C_{AB}$ & $C_{AC}$ & $C_{AD}$ & $C_{BC}$ & $C_{BD}$ & $C_{CD}$ & $C_{GM}$ \\
\midrule
$C_1$ & 0 & 0 & 0 & 0 & 0 & 0 & 0 & 0 & 0 & 0 & 0 \\
$C_3$ & 0 & 0 & 0 & 0 & 0 & $\neq 0$ & 0 & 0 & 0 & 0 & 0 \\
$C_9$ & $\neq 0$ & $\neq 0$ & 0 & 0 & 0 & $\neq 0$ & 0 & 0 & $\neq 0$ & 0 & 0 \\
$C_{11}$ & 0 & 0 & 0 & 0 & $\neq 0$ & $\neq 0$ & 0 & $\neq 0$ & 0 & 0 & 0 \\
$C_{15}$ & 0 & 0 & 0 & 0 & 0 & 0 & 0 & 0 & 0 & 0 & 0 \\
$C_{19}$ & 0 & 0 & 0 & 0 & $\neq 0$ & $\neq 0$ & $\neq 0$ & $\neq 0$ & $\neq 0$ & $\neq 0$ & $\neq 0$ \\
$C_{22}$ & 0 & 0 & 0 & 0 & 0 & 0 & 0 & 0 & $\neq 0$ & 0 & $\neq 0$ \\
$C_{26}$ & $\neq 0$ & 0 & 0 & 0 & 0 & 0 & 0 & 0 & 0 & 0 & $\neq 0$ \\
$C_{28}$ & $\neq 0$ & $\neq 0$ & 0 & 0 & $\neq 0$ & $\neq 0$ & $\neq 0$ & $\neq 0$ & $\neq 0$ & $\neq 0$ & $\neq 0$ \\
$C_{31}$ & $\neq 0$ & $\neq 0$ & 0 & $\neq 0$ & $\neq 0$ & 0 & $\neq 0$ & $\neq 0$ & 0 & $\neq 0$ & $\neq 0$ \\
$C_{33}$ & $\neq 0$ & $\neq 0$ & $\neq 0$ & $\neq 0$ & 0 & $\neq 0$ & $\neq 0$ & $\neq 0$ & $\neq 0$ & 0 & $\neq 0$ \\
$C_{36}$ & 0 & 0 & 0 & 0 & 0 & 0 & $\neq 0$ & 0 & $\neq 0$ & 0 & $\neq 0$ \\
$C_{39}$ & $\neq 0$ & 0 & 0 & 0 & 0 & $\neq 0$ & 0 & 0 & $\neq 0$ & 0 & $\neq 0$ \\
$C_{43}$ & $\neq 0$ & $\neq 0$ & 0 & 0 & 0 & $\neq 0$ & 0 & 0 & $\neq 0$ & 0 & $\neq 0$ \\
$C_{48}$ & 0 & $\neq 0$ & 0 & 0 & 0 & 0 & 0 & 0 & 0 & 0 & $\neq 0$ \\
$C_{50}$ & 0 & 0 & 0 & 0 & 0 & 0 & 0 & 0 & 0 & 0 & $\neq 0$ \\
$C_{57}$ & $\neq 0$ & 0 & 0 & 0 & $\neq 0$ & $\neq 0$ & 0 & 0 & $\neq 0$ & $\neq 0$ & $\neq 0$ \\
$C_{60}$ & $\neq 0$ & 0 & 0 & $\neq 0$ & $\neq 0$ & $\neq 0$ & $\neq 0$ & $\neq 0$ & $\neq 0$ & $\neq 0$ & $\neq 0$ \\
$C_{62}$ & 0 & $\neq 0$ & 0 & 0 & $\neq 0$ & 0 & $\neq 0$ & $\neq 0$ & 0 & $\neq  0$ & $\neq 0$ \\
$C_{65}$ & $\neq 0$ & $\neq 0$ & 0 & $\neq 0$ & $\neq 0$ & $\neq 0$ & $\neq 0$ & $\neq 0$ & $\neq 0$ & $\neq 0$ & $\neq 0$ \\
$C_{67}$ & $\neq 0$ & $\neq 0$ & $\neq 0$ & 0 & 0 & $\neq 0$ & 0 & 0 & $\neq 0$ & 0 & $\neq 0$ \\
$C_{68}$ & $\neq 0$ & 0 & 0 & 0 & $\neq 0$ & $\neq 0$ & 0 & $\neq 0$ & 0 & 0 & $\neq 0$ \\
$C_{72}$ & 0 & 0 & 0 & $\neq 0$ & 0 & 0 & 0 & 0 & 0 & 0 & $\neq 0$ \\
$C_{75}$ & 0 & $\neq 0$ & 0 & 0 & 0 & $\neq 0$ & $\neq 0$ & 0 & 0 & $\neq 0$ & $\neq 0$ \\
$C_{80}$ & 0 & $\neq 0$ & 0 & $\neq 0$ & 0 & $\neq 0$ & 0 & 0 & 0 & 0 & $\neq 0$ \\
$C_{82}$ & $\neq 0$ & $\neq 0$ & $\neq 0$ & $\neq 0$ & $\neq 0$ & $\neq 0$ & 0 & 0 & $\neq 0$ & $\neq 0$ & $\neq 0$ \\
\bottomrule
\end{tabular}
\captionof{table}{Classification of entanglement classes by polynomial invariants, bipartite concurrences and general multipartite concurrence \cite{JafarizadehPhysLettA2019}.}\label{table2}
\end{table}

Finally, we consider the single qubit averaged von Neumann entropy:
\begin{eqnarray}
    \bar{S}(\ket{\psi}) = -\frac{1}{N}\sum_j \text{Tr}(\rho_j\log\rho_j),
\end{eqnarray}
where $\rho_j = \mbox{Tr}_{\bar{j}}(\ket{\psi}\bra{\psi})$ is the $j$th qubit reduction and $\bar{j}$ denotes the complementary set of $\{j\}$ within the set $\{1,\dots,N\}$, for $N$ qubit systems.

Next subsection introduces another fundamental ingredient of this work: the qubit-ADAPT-VQE algorithm.

\subsection{Qubit-ADAPT-VQE}
The Variational Quantum Eigensolver \cite{PeruzzoNatCommun2014} (VQE) is a hybrid quantum-classical optimization algorithm used to find the ground state energy of a Hamiltonian $H_f$ given a parametrized ansatz $\ket{\psi(\bm{\theta})}$, that is:
\begin{eqnarray}
    E = \min_{\bm{\theta}} \bra{\psi(\bm{\theta})}\hat{
    H_f\ket{\psi(\bm{\theta})}
    }.
\end{eqnarray}
Here, the variational principle guarantees that $E$ is an upper bound for the minimal energy of the Hamiltonian. Such upper bound implies a good approximation of the global minimum provided that the ansatz $\ket{\psi(\bm{\theta})}$ is capable of getting arbitrarily close to the true ground state of $H_f$.

VQE has been widely applied to molecular systems and its performance strongly depends on the choice of the  ansatz. Common fixed ansatz based on excitation operators often lead to large circuit depth and suffers from barren plateaus \cite{MaoCommunPhys2024}. In this context, the ADAPT-VQE algorithm was introduced for building a suitable ansatz in an iterative and adaptive way, yielding more compact circuits and  improving robustness against barren plateaus, thus enhancing its potential for near-term quantum applications \cite{GrimsleyNatCommun2019, GrimsleynpjQuantumInf2023}. 

Our work is based on a subsequent variant of the ADAPT-VQE, so called qubit-ADAPT-VQE \cite{LunTangPRXQuantum2021}, where the operator pool  consists of Pauli string operators. That is, the operator pool consists of a set of Pauli strings $\{A_k\}_{k=1}^M$, where each $A_k \in \{X, Y, Z, I\}^{\otimes N}$, and $M$ denotes the total number of operators in the pool. Here $X$, $Y$ and $Z$ are the Pauli operators acting on one qubit, and $I$ is the identity operator. The algorithm is initialized with a convenient reference state $\ket{\psi_0}$ and iteration count $j=0$. For each operator $A_k$ in the pool, the energy gradient with respect to a variational parameter $\theta_k$ is evaluated when applying the unitary $e^{-i\theta_k A_k}$ to the $j$-th ansatz state $\ket{\psi_j}$, at $\theta_k = 0$. This derivative takes the form of a commutator expectation value:

\begin{eqnarray}
g_k = \left. \frac{\partial E}{\partial \theta_k} \right|_{\theta_k = 0} = i\bra{\psi_j} [A_k,H] \ket{\psi_j},
\end{eqnarray}
 and the collection of all such values defines the gradient vector $\vec{g} = (g_1, \dots, g_M)$. If the 2-norm $\|\vec{g}\|_2$ is less than a predefined threshold $\epsilon$, the algorithm finds an optimal ansatz and executes a final VQE subroutine. Otherwise, the operator $A_k$ corresponding to the maximum value  $|g_k|$ is selected (denoted $A^{\text{max}}$) and added to the ansatz. The ansatz is then updated as:

\begin{eqnarray}\label{ansatz}
\ket{\psi_{j+1}} = e^{-i\theta_j A^{\text{max}}} \ket{\psi_j},
\end{eqnarray}
\noindent
where $\theta_j$ is an additional parameter. Based on the constructed ansatz (\ref{ansatz}), a VQE optimization is then performed to update all parameters to the optimum value, obtaining an optimal state. This state is then set as the input for the next iteration count of the algorithm and the process repeats until $\|\vec{g}\|_2<\epsilon$.
The authors in \cite{LunTangPRXQuantum2021} have proposed the use of reduced operator pools to simplify the ansatz construction in qubit-ADAPT-VQE. However, we are interested in the entanglement generation; therefore we use the complete set of Pauli strings for four-qubit. Its operator pool, composed of Pauli strings, provides a more versatile structure that can be applied to any system admitting a qubit-based representation. 
\subsection{Spin models: XY and Heisenberg XXZ }
Understanding how entanglement emerges and is captured by variational quantum circuits requires to consider Hamiltonians where quantum correlations are relevant. In this context, spin chains provide a natural testing models, as they exhibit a wide range of quantum phases and entanglement patterns while remaining amenable to analytical and numerical treatment. Among these, the XY and Heisenberg XXZ models are particularly compelling choices. Both are paradigmatic in the study of quantum magnetism and the entanglement properties of their ground states have been deeply analyzed \cite{LatorreQuantumInfComput2004,AmicoRevModPhys2008}. 

The XY model, which includes the transverse-field Ising chain as a special case, undergoes a quantum phase transition that leads to strong entanglement \cite{OsbornePhysRevA2002}. This makes it specially relevant for assessing the ability of variational algorithms to capture critical behavior and long-range quantum correlations. The XY model has associated the following Hamiltonian:
\begin{eqnarray}\label{XY}
    H_{XY} = \sum_{j=1}^N\left( \frac{1+\gamma}{2}X_jX_{j+1} + \frac{1-\gamma}{2}Y_jY_{j+1} + \lambda Z_j \right),
\end{eqnarray}
where $\gamma$ is the anisotropy parameter in the X-Y plane and periodic boundary conditions were imposed in subindices, identifying site $N+1$ with site $1$.

The XXZ model, on the other hand, exhibits a broader variety of quantum phases depending on the anisotropy parameter $\Delta$ in z-axis, with ground states that display multipartite entanglement even near critical regimes \cite{GuPhysRevA2005, StasinskaPhysRevA2014}. The Hamiltonian for one-dimensional chain is given by:
\begin{eqnarray}\label{XXZ}
    H_{XXZ} = \sum_{j=1}^N\left( J[X_jX_{j+1} + Y_jY_{j+1}] +\Delta Z_jZ_{j+1} + \lambda Z_j \right),
\end{eqnarray}
where the same periodic conditions apply to indices. By chosing XY and XXZ models, we ensure that our benchmarks probe nontrivial entanglement generation, providing a test for the ability of the qubit-ADAPT-VQE to identify highly entangled ground states.

\section{Results}\label{results}
In this section, we present the main results on entanglement generation by using the algebraic classification framework for qubit-ADAPT-VQE. Specifically, we identify the different entanglement classes that arise in the spin models and discuss their entanglement properties. Finally, we evaluate the performance of qubit-ADAPT-VQE under these considerations for different scenarios of initial states.  

\subsection{Entanglement classification in ground states of spin models}
We start our study by computing the exact ground states of the XY and XXZ spin models and classifying them according to the entanglement structure induced by the invariants introduced in Table \ref{table2}. Our goal is to identify the different entanglement classes that naturally emerge in these models across various anisotropy parameters and focus on them for ground state simulation with the qubit-ADAPT-VQE algorithm.
\begin{figure}[H]
    \centering
    \includegraphics[width=0.9\linewidth]{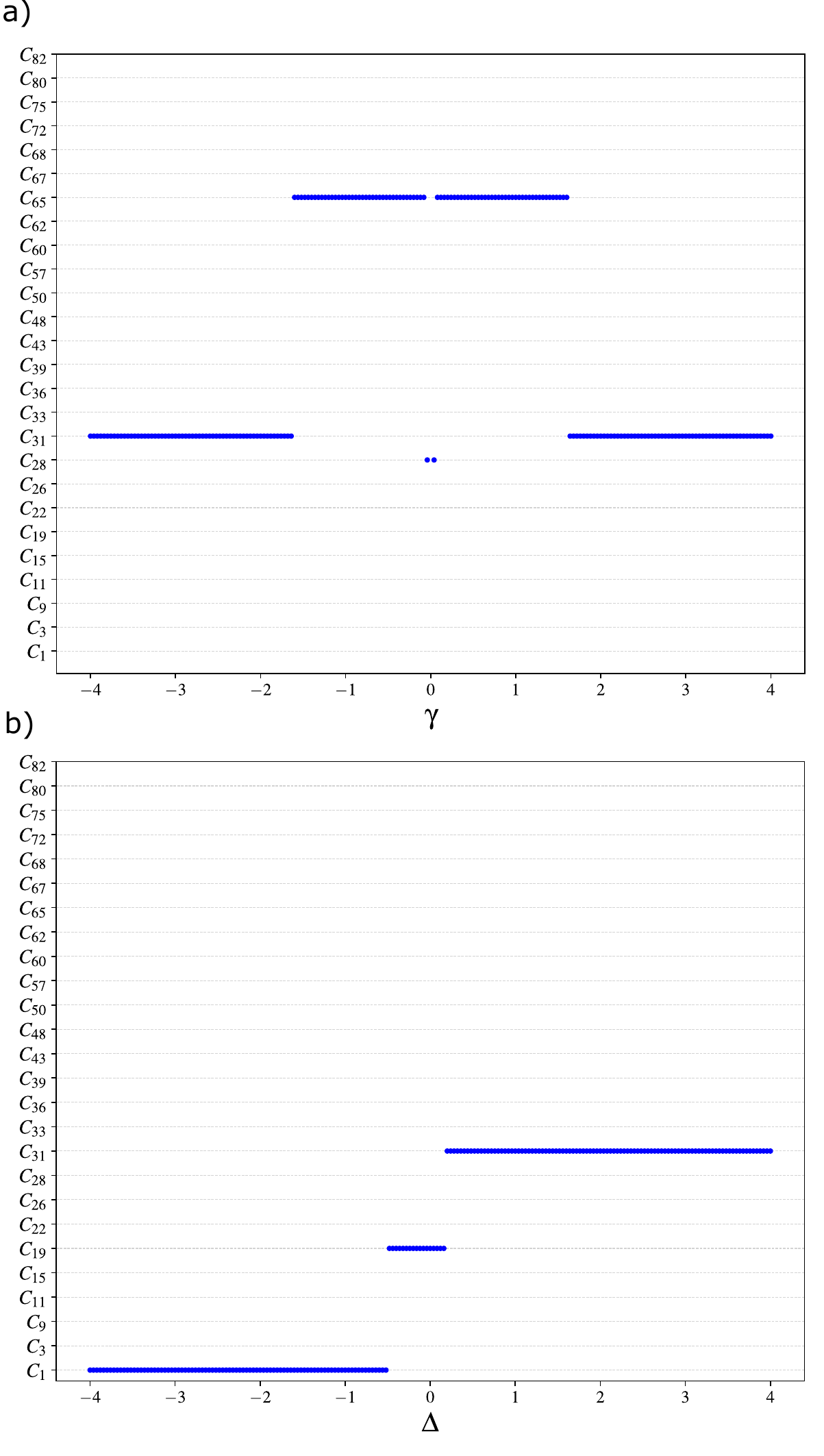}
    \caption{Entanglement classes of the exact ground state in spin models. a) XY model, defined in (\ref{XY}), as a function of the anisotropy parameter $\gamma$, with $\lambda = 1$. b) XXZ model, defined in (\ref{XXZ}), as a function of the anisotropy parameter $\Delta$, with $\lambda = 1$ and $J = 1$.}
    \label{fig01} 
\end{figure}
For the XY model, Fig.~\ref{fig01} a) shows the vertical axis labeled with the 26 entanglement classes (excluding $C_0$) and their occurrence for $\lambda = 1$ and different values of the anisotropy parameter $\gamma$. Three entangled classes are observed, exhibiting symmetric behavior around $\gamma = 0$. As $\gamma$ increases from 0, entanglement in the ground state emerges. The first to appear is class $C_{28}$, characterized by nonzero bipartite concurrences between all qubit pairs and a finite GM entanglement. As $\gamma$ increases further, class $C_{65}$ emerges, exhibiting the same concurrence and GM features as $C_{28}$ but differing in its set of polynomial invariants. Finally, class $C_{31}$ appears, in which two pairs of qubits are disentangled while the GM entanglement remains nonzero. The same behavior is appreciated for negative values of $\gamma$. 

For the XXZ model, the quantum phase transition occurs at $\Delta = -0.5$ with $\lambda = 1$ and $J = 1$. As $\Delta$ increases, class $C_{19}$ emerges, which, similar to the previous case, is characterized by nonzero concurrences between all qubit pairs and a finite GM entanglement, but differs with $C_{28}$ in its set of polynomial invariants. Further increasing $\Delta$ leads to the appearance of class $C_{31}$. 

Note that for large values of $\gamma$ and $\Delta$, both models exhibit the same entanglement structure, corresponding to class $C_{31}$. However, the specific values of the entanglement measures can differ between the two models, even when their ground states belong to the same entanglement class. For the purpose of testing the qubit-ADAPT-VQE algorithm, we focus on cases where distinct entanglement classes appear in the ground state. For the XY model, we consider $\gamma = 1$ and $\gamma = 3$, corresponding to classes $C_{65}$ and $C_{31}$, respectively, while for the XXZ model we select $\Delta = -0.1$ and $\Delta = 3$, corresponding to classes $C_{19}$ and $C_{31}$, respectively.\\

\subsection{Entangled ground state simulation with qubit-ADAPT-VQE}

For the qubit-ADAPT-VQE and for several variational quantum algorithms, the choice of the initial state plays a significant role in its performance \cite{LaroccaNatRevPhys2025}. In molecular simulations, the initial state is typically the Hartree-Fock state. In the present case, however, we do not have an analogous reference state informed by the specific system under study. Instead, we consider the choice of the initial state as a means to investigate performance of the qubit‑ADAPT‑VQE across different scenarios. 

In further algorithms such as QAOA, the initial state is often chosen as the uniform superposition over the computational basis, primarily because it assigns the same probability to every basis state, thus starting without a bias. The uniform superposition is given by the fully separable state
\begin{eqnarray}
    \ket{s} = \frac{1}{\sqrt{2}} \left( \ket{0} + \ket{1} \right)^{\otimes N},
\end{eqnarray}
where in our case $N=4$. By starting from this state, we ensure that any reduction in energy and the generation of entanglement arise entirely during the optimization process. We do the simulation of qubit-ADAPT-VQE with the initial state $\ket{\psi_0} = \ket{s}$.

Another interesting case arises when the initial condition is an entangled state. This tests flexibility of the algorithm  when modifying entanglement structure of the state, while simultaneously minimizing the energy by selecting appropriate operators from the operator pool. Therefore we also implement simulation of the qubit-ADAPT-VQE with the initial states
\begin{eqnarray}
    \ket{\psi_0} = \ket{C_l}
\end{eqnarray}
where $l$ denotes the label corresponding to one of the 27 entanglement classes, and the state $\ket{C_l}$ is the representative of class $l$, as listed in Table~\ref{table1}. Simulations were performed for each of the 27 classes. Table~\ref{table_l} in Appendix \ref{appendix2} summarizes the percentage error and iteration counts required by qubit-ADAPT-VQE when initialized with representative states from each of the entanglement classes.

To illustrate the performance of the algorithm, we show the energy obtained at iteration $j$ of the qubit-ADAPT-VQE:
\begin{eqnarray}
    E^{j} = \bra{\psi_j} H_{\alpha} \ket{\psi_j},
\end{eqnarray}
where $\alpha = XY, XXZ$ depending on the chosen model. 
\begin{figure}[H]
    \centering    
    \includegraphics[width=0.9\linewidth]{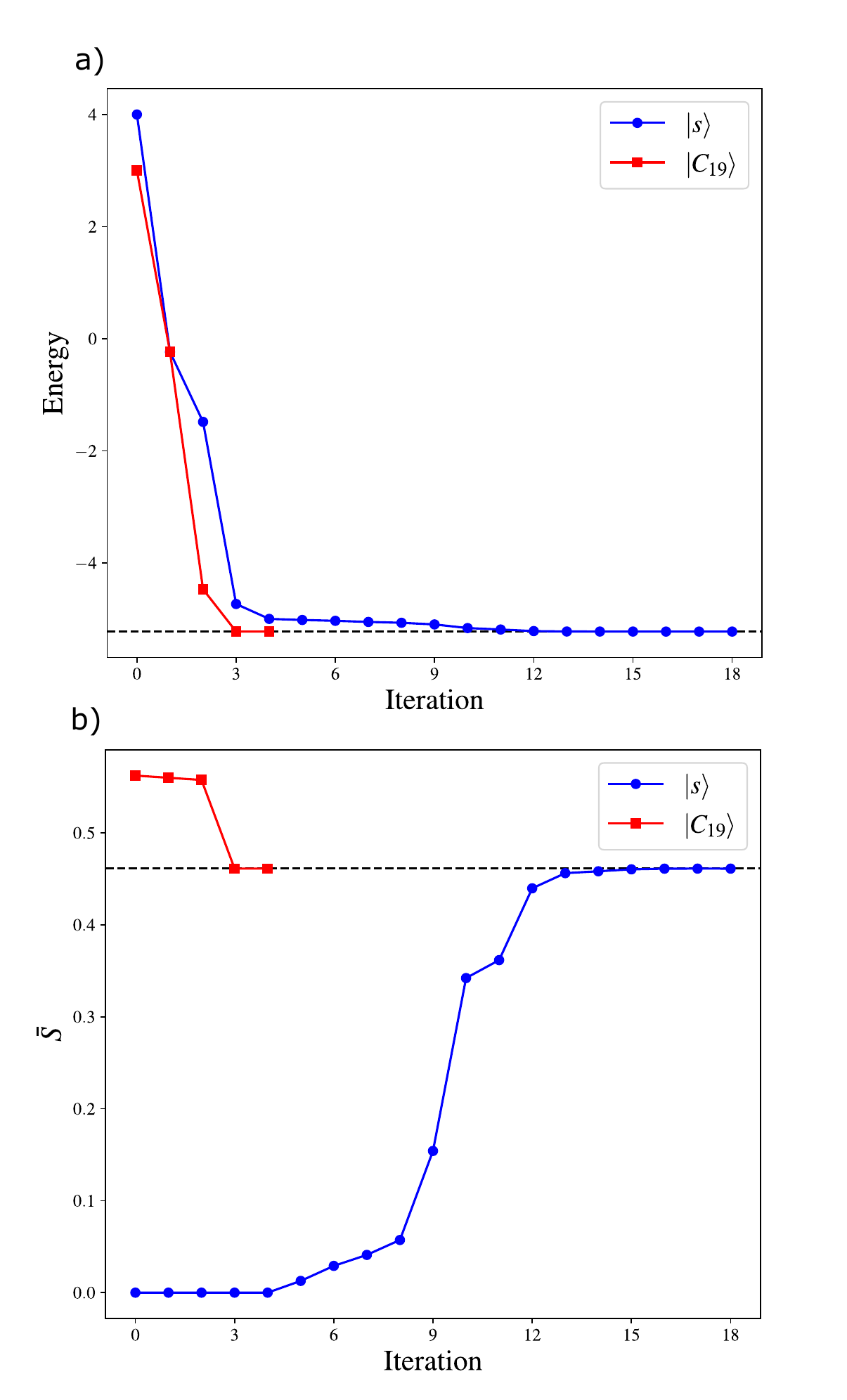}
    \caption{Ground state simulation for the XY model with $\gamma = 1$ (a) Energy $E^{j}$ at iteration $j$ of the qubit-ADAPT-VQE. (b) Average von Neumann entropy $\bar{S}$ at iteration $j$. In subfigures a) and b) dashed lines represent the exact energy and exact averaged entropy of the ground state, respectively.}
    \label{fig02} 
\end{figure}

Fig.~\ref{fig02} a) shows $E^j$ when the algorithm is initialized with $\lvert s\rangle$ and with $\lvert C_{19}\rangle$, the latter being the best-performing representative state among all entanglement classes, for the XY model with $\gamma= 1$ and $\lambda = 1$. We observe a significant reduction in the number of iterations, with the entangled state $\ket{C_{19}}$ reaching convergence in only 4 iterations, compared to 18 iterations required by the separable state $\ket{s}$. In Fig.~\ref{fig02} b), we present the average Von Neumann entropy $\bar{S}$ of the state at each iteration. At iteration $j = 0$, the initial entanglement of $\ket{C_{19}}$ is already nonzero, as expected, but also closer to the target entanglement value indicated by the dashed black lines. It is worth noting that the representative state from the entanglement class to which the ground state truly belongs does not always yield the best performance in terms of energy error and iteration count, as shown in Table~\ref{table_l}. This is because the representative state may not necessarily provide the lowest initial energy within all elements of its class. Since the energy gradient is the criterion used to select operators from the pool, the initial energy plays an important role as well as the initial amount of entanglement. Furthermore, certain initial conditions may lead to small cost function gradients in the vicinity of a given iteration, and the optimization procedure might be trapped in regions where the energy change is smaller, as we can observe in Fig.~\ref{fig02} for $\ket{s}$, at the lastest iterations. We explore this aspect in more detail below.
\begin{figure}[H]
    \centering    
    \includegraphics[width=0.9\linewidth]{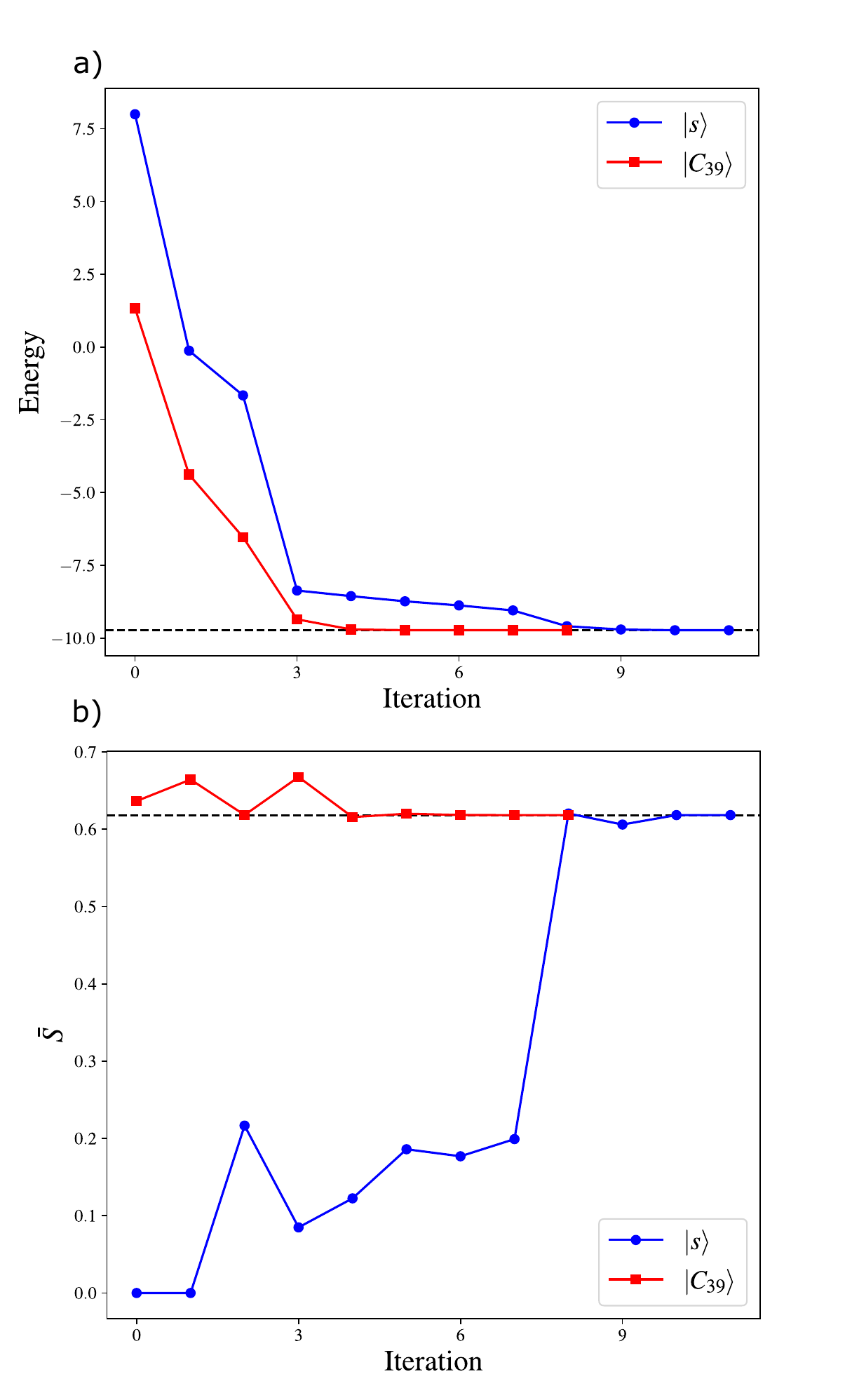}
    \caption{Ground state simulation of XY model with $\gamma = 3$ (a) Energy $E^{j}$ at iteration $j$ of the qubit-ADAPT-VQE. (b) Average von Neumann entropy $\bar{S}$ at iteration $j$. In both cases, the dashed line represents the energy and $\bar{S}$ of the exact ground state.}
    \label{fig03} 
\end{figure}

In Fig.~\ref{fig03} we shows the energy $E^{j}$ and averaged entropy $\bar{S}$ for the XY model, when considering $\gamma = 3$ and $\lambda = 1$ for the initial states $\ket{s}$. In such a case, the best performing representative of entanglement classes is $\ket{C_{39}}$. We observe that the initial entanglement $\bar{S}$ of $\ket{C_{39}}$ is notably closer to the target entanglement, and its initial energy is lower than the one associated to the separable state $\ket{s}$, leading to an improved performance. However, we also observe from iteration $j=4$ onward that the algorithm becomes trapped in a region where the cost function gradients remain small in a neighborhood. Despite this, it eventually converges after a further optimization at iteration $j=8$.
\begin{figure}[H]
    \centering    
    \includegraphics[width=0.9\linewidth]{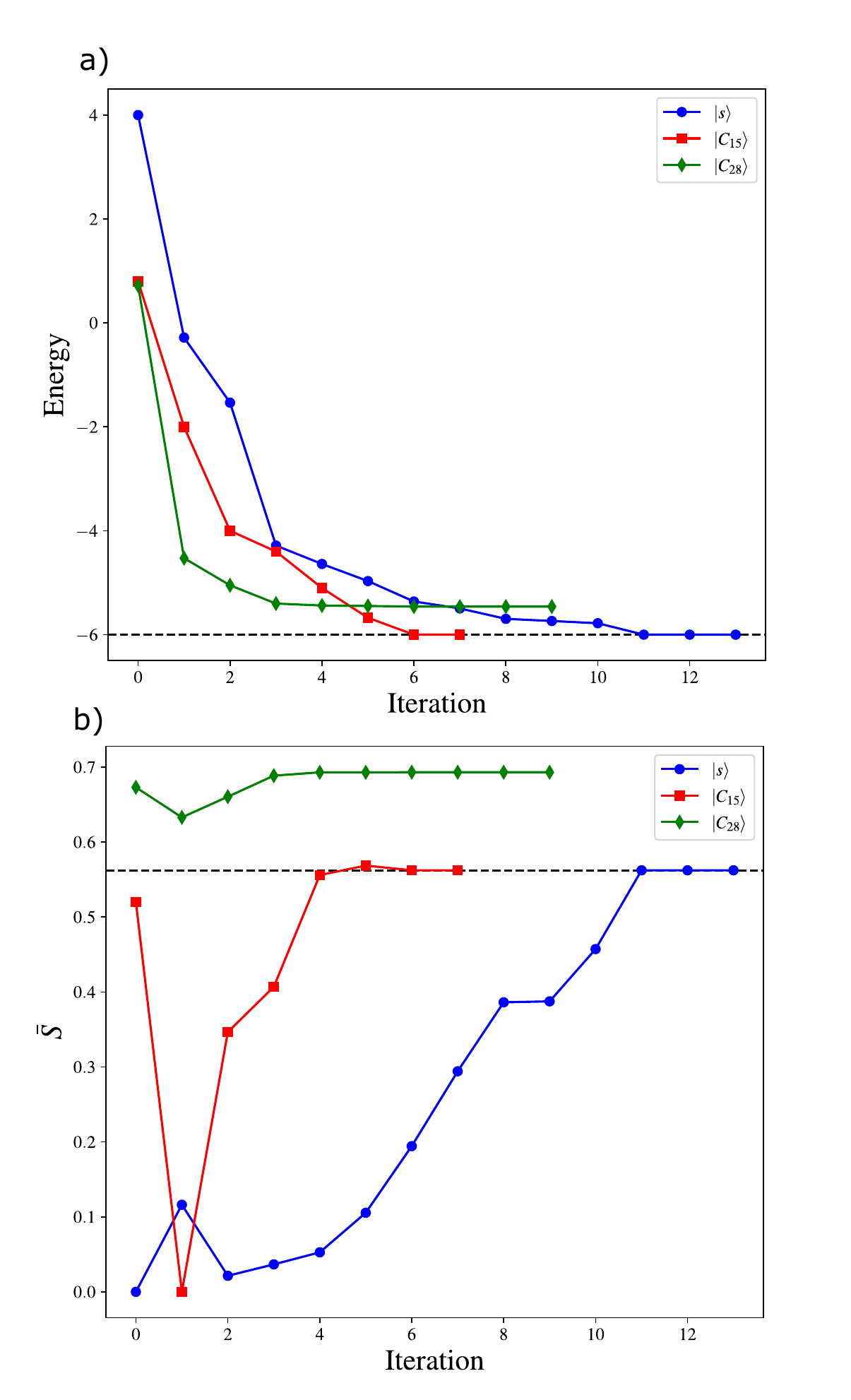}
    \caption{Ground state simulation of XXZ model with $\Delta = -0.1$ (a) Energy $E^{j}$ at iteration $i$ of the qubit-ADAPT-VQE. (b) Average von Neumann entropy $\bar{S}$ at iteration $j$. In both cases, the dashed line represents the energy and $\bar{S}$ of the exact ground state.}
    \label{fig04} 
\end{figure}
In Fig.~\ref{fig04}, we present the energy $E^{j}$ and averaged entropy  $\bar{S}$ as a function of the number of iterations $j$ for the XXZ model, with $\Delta = -0.1$ and $\lambda = J = 1$. We compare the performance starting from the separable state $\ket{s}$, the representative from the entanglement classes that yields the best result, $\ket{C_{15}}$, and an additional case, $\ket{C_{28}}$, whose initial energy is similarly close to the one induced by $\ket{C_{15}}$. Interestingly, the evolution from $\ket{C_{28}}$ does not converge to the true ground state, despite its comparable initial energy. Numerical simulations indicate that the algorithm instead converges to the first excited state. This behavior arises because the entanglement of the first excited state is closer to that of the initial state $\ket{C_{28}}$, and the gradients of the cost function are small in the neighborhood of the preceding iteration from $j=4$ onward. Therefore, in this case, the initial energy plays a less significant role in guiding convergence.

\begin{figure}[H]
    \centering    
    \includegraphics[width=0.9\linewidth]{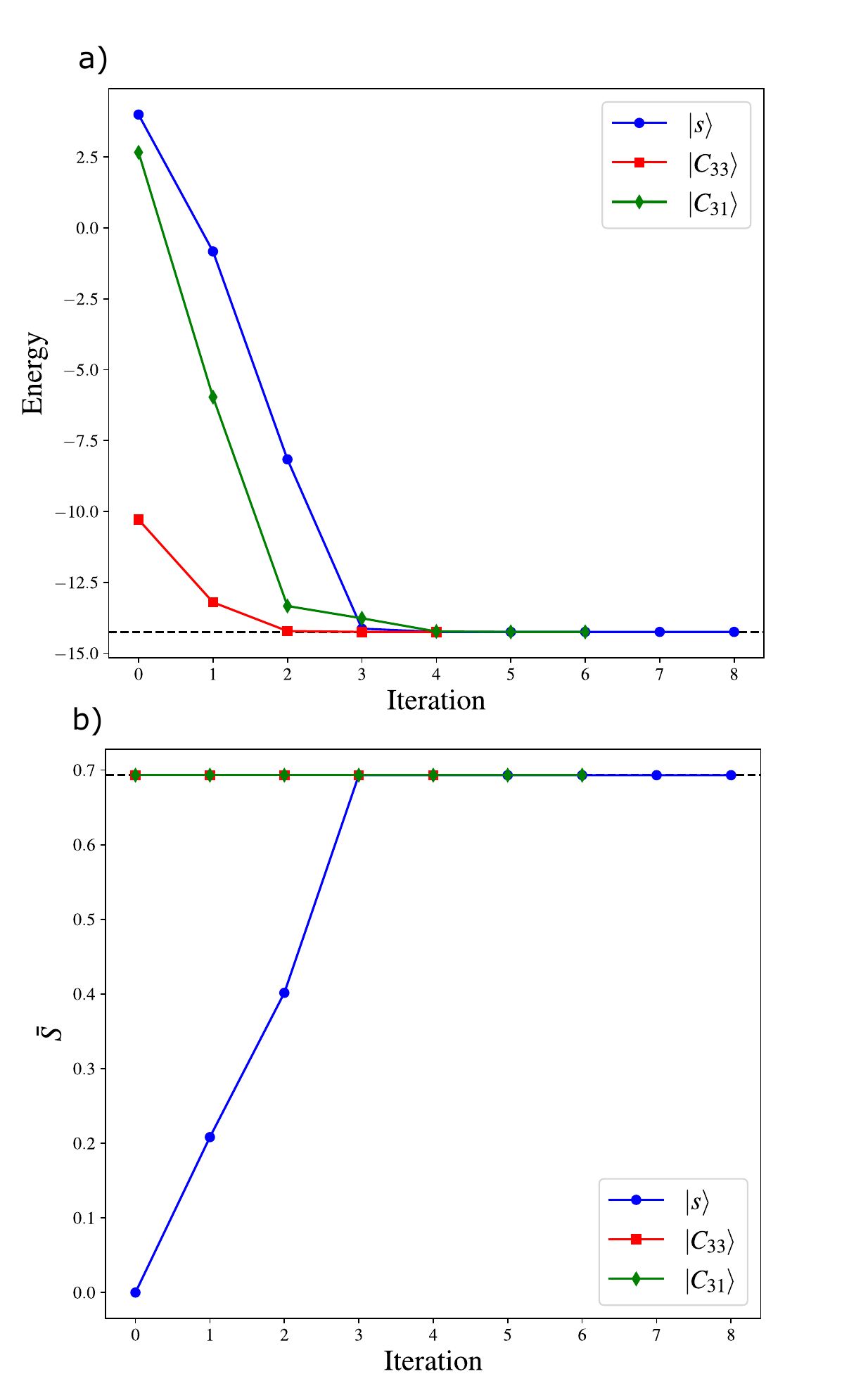}
    \caption{Ground state simulation of XXZ model with $\Delta = 3$ (a) Energy $E^{j}$ at iteration $j$ of the qubit-ADAPT-VQE. (b) Average von Neumann entropy $\bar{S}$ at iteration $j$. In both cases, the dashed line represents the energy $E^{j}$ and averaged entropy $\bar{S}$ of the exact ground state.}
    \label{fig05} 
\end{figure}

In Fig.~\ref{fig05} we show plots similar to Fig. ~\ref{fig04} but for the XXZ model, considering $\Delta = 3$. We show the initial states $\ket{s}$, the best performing representative state $\ket{C_{33}}$ and another representative state with the same entanglement $\bar{S}$, that is the state $\ket{C_{31}}$. In this case, the initial state $\ket{C_{31}}$ converges to the correct ground state after 6 iterations, a smaller number of iterations than those required by the separable state $\ket{s}$ and higher than those required by the state $\ket{C_{33}}$. We observe that both representative states exhibit the same average entanglement $\bar{S}$ as the exact ground state, yet differ in their initial energies, which influences the number of iterations required for convergence.
However, as shown in Figs. \ref{fig02} a) and \ref{fig04} a), there does not to be a one-to-one correspondence between the initial energy, entanglement and the success of convergence to the ground state. Overall, although initial entanglement values close to those of the ground state generally correlate with faster convergence, the proximity of the initial energy and the emergence of small cost function gradients in the vicinity of the previous iteration appear to be more decisive factors. It is also worth noting that, despite the additional iterations required when such flat regions arise, convergence is ultimately achieved in most cases, with very low percentage error, as shown in Table~\ref{table_l}. As a further observation, adding extra operators to the ansatz does not justify the associated increase of the circuit depth, since the resulting improvement in energy is negligible. 
\begin{figure}[H]
    \centering    
    \includegraphics[width=0.9\linewidth]{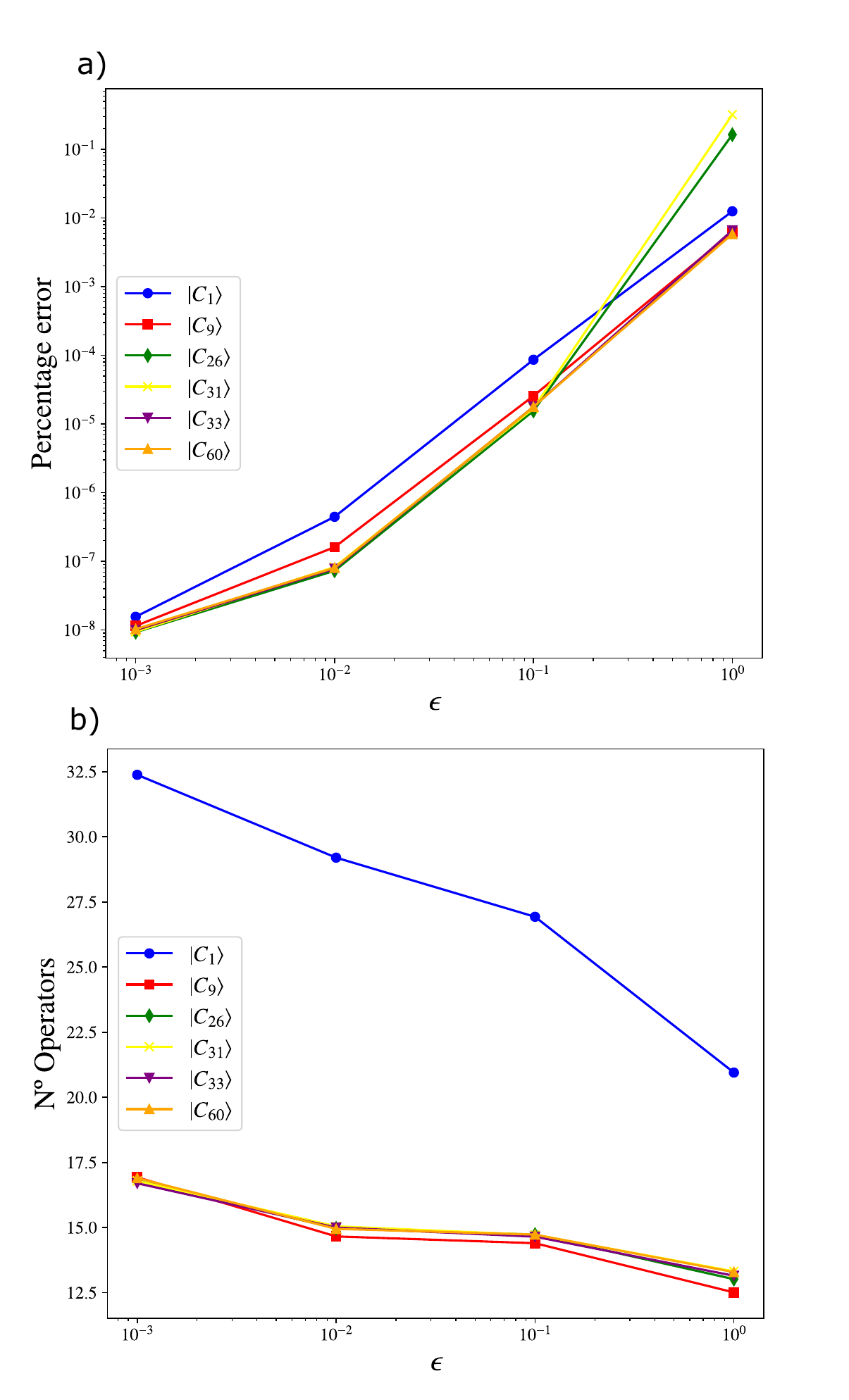}
    \caption{Averaged results over 100 LU equivalent states realizations for the classes 
$C_1, C_9, C_{26}, C_{31}, C_{33}$ and $C_{60}$ for different values of the threshold $\epsilon$. 
a) Average percentage error as a function of $\epsilon$. 
b) Average number of operators appended to the ansatz as a function of $\epsilon$.
}
    \label{fig06} 
\end{figure}
In this work, we focus on the performance of the algorithm and on its ability to converge under its own selection criterion, without considering the resource requirements for a realistic implementation, as our analysis is intended as a theoretical study of the algorithm. The most unfavorable performance of the representative states occurs in the XXZ case with $\Delta = -0.1$, where twelve of them exceed one percentage error, as shown in Table \ref{table_l}. One might suspect that this behavior is linked to the entanglement structure of the specific class, but the next section shows that this is not the case.

\subsection{Local unitary equivalent initial states}
We can explore different initial energy while preserving the entanglement structure of the representative states by applying local unitary (LU) transformations. This allows us to probe different regions of the energy landscape, including those in which small cost function gradients may occur. The initial state after the LU transformation is given by:
\begin{eqnarray}
    \ket{\psi_0} = (U_A \otimes U_B \otimes U_C \otimes  U_D) \ket{C_l}
\end{eqnarray}
 where $[U_A,U_B,U_C,U_D]$ are local randomly generated by considering the Haar measure distribution. We considered a set of 100 different cases of $[U_A,U_B,U_C,U_D]$ for every entanglement class. Table \ref{table_100} shows the results for the average percentage error of the final energy of 100 cases, with average number of operators added to ansatz. We observe that, in contrast to Table~\ref{table_l}, all entanglement classes lead to highly accurate solutions, with percentage errors below $10^{-5}$ in nearly all cases. Notably, the separable class successfully converges to the minimum with low percentage error, demonstrating effective entanglement generation alongside energy minimization across all the considered models. Moreover, the entangled state classes also achieve convergence to the minimum, even when the class does not match the one of the exact ground state. 

\begin{figure}[H]
    \centering    
    \includegraphics[width=0.99\linewidth]{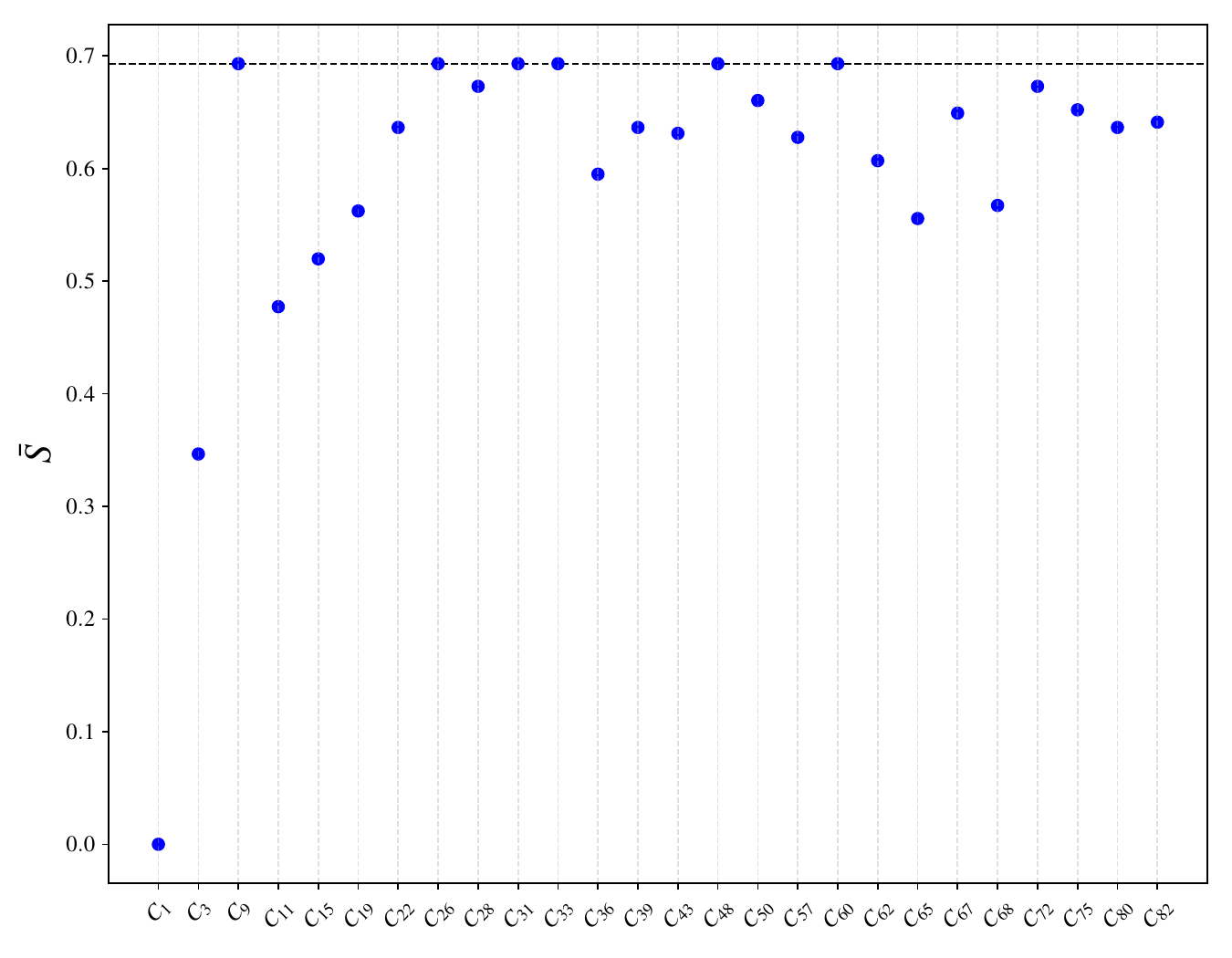}
    \caption{Average von Neumann entropy $\bar{S}$ of the representative states of all 27 entanglement classes. The dashed line correspond to maximmum value possible for $\bar{S}$, that is $\ln(2)$. }
    \label{fig07} 
\end{figure}

Regarding the average number of operators added to the ansatz, some classes exhibit a clear advantage by requiring a lower number of operators on average. The classes $C_9, C_{26}, C_{31}, C_{33}$ and $C_{60}$ show this trend consistently across the four spin models considered, with the effect being particularly pronounced in the XXZ model with $\Delta = 3$, as reported in Table~\ref{table_100}. To examine this behavior in greater detail, we again consider 100 LU equivalent states for these classes and explore different values of the threshold $\epsilon$, since this parameter influences the number of operators appended to the ansatz. Figure~\ref{fig06} shows the results for the XXZ model with $\Delta = 3$, including the separable class $C_1$. Although no substantial differences in average percentage error are observed, the advantage of these classes, expressed in the lower number of operators required, persists across the entire range of threshold values examined. These classes include states with entanglement but no GM entanglement ($C_9$), states with GM entanglement only ($C_{26}$), states with both pairwise entanglement and GM entanglement ($C_{31}$ and $C_{33}$), and states with all pairs entangled together with GM entanglement ($C_{60}$). The only common feature in their invariants is that $h_1 \neq 0$. 
We also evaluate the general measure of entanglement $\bar{S}$ for each representative state of the classes. It is important to note that this quantity is invariant under LU transformations. As shown in Fig.~\ref{fig07}, the classes with the largest values of $\bar{S}$ are $C_9, C_{26}, C_{31}, C_{33}, C_{60}$, and $C_{48}$. This helps explain their faster convergence to the solution and the lower number of operators required in the ansatz. However, although class $C_{48}$ exhibits the highest value of $\bar{S}$, it does not show a clear advantage in terms of operator added. This class satisfies $h_1 = 0$, in contrast with $C_9, C_{26}, C_{31}, C_{33}$, and $C_{60}$, indicating that the general entanglement measure alone $\bar{S}$ is not sufficient and that the entanglement class also plays a relevant role.

\begin{widetext}
\centering
\begin{tabular}{|c|c|c|c|c|}
\hline
\textbf{Class} & 
\textbf{XY, $\gamma=1$ (N° Ops)} & 
\textbf{XY, $\gamma=3$ (N° Ops)} & 
\textbf{XXZ, $\Delta = -0.1$ (N° Ops)} & 
\textbf{XXZ, $\Delta = 3$ (N° Ops)} \\
\hline
$C_1$  & $3.67\times10^{-6}$ (29) & $1.06\times10^{-6}$ (29) & $2.60\times10^{-6}$ (30) & $4.45\times10^{-7}$ (29) \\
\hline
$C_3$  & $3.06\times10^{-6}$ (29) & $8.09\times10^{-7}$ (29) & $2.73\times10^{-6}$ (29) & $7.17\times10^{-7}$ (29) \\
\hline
$C_9$  & $5.28\times10^{-6}$ (28) & $1.70\times10^{-6}$ (28) & $2.70\times10^{-1}$ (26) & $1.61\times10^{-7}$ (15) \\
\hline
$C_{11}$ & $2.27\times10^{-6}$ (30) & $1.00\times10^{-6}$ (29) & $2.61\times10^{-6}$ (30) & $2.95\times10^{-7}$ (30) \\
\hline
$C_{15}$ & $2.83\times10^{-6}$ (30) & $7.32\times10^{-7}$ (30) & $2.68\times10^{-6}$ (29) & $4.50\times10^{-7}$ (29) \\
\hline
$C_{19}$ & $2.17\times10^{-6}$ (30) & $8.43\times10^{-7}$ (30) & $2.19\times10^{-6}$ (30) & $4.04\times10^{-7}$ (30) \\
\hline
$C_{22}$ & $3.64\times10^{-6}$ (30) & $6.38\times10^{-7}$ (30) & $2.97\times10^{-6}$ (29) & $2.83\times10^{-7}$ (30) \\
\hline
$C_{26}$ & $3.73\times10^{-6}$ (27) & $1.18\times10^{-6}$ (29) & $8.99\times10^{-2}$ (26) & $7.29\times10^{-8}$ (15) \\
\hline
$C_{28}$ & $2.34\times10^{-6}$ (30) & $8.00\times10^{-7}$ (30) & $2.81\times10^{-6}$ (30) & $2.60\times10^{-7}$ (30) \\
\hline
$C_{31}$ & $3.92\times10^{-6}$ (25) & $1.40\times10^{-6}$ (29) & $2.70\times10^{-1}$ (24) & $8.26\times10^{-8}$ (15) \\
\hline
$C_{33}$ & $2.85\times10^{-6}$ (27) & $1.85\times10^{-6}$ (29) & $1.80\times10^{-1}$ (26) & $7.80\times10^{-8}$ (15) \\
\hline
$C_{36}$ & $3.83\times10^{-6}$ (30) & $1.04\times10^{-6}$ (29) & $3.19\times10^{-6}$ (30) & $4.20\times10^{-7}$ (30) \\
\hline
$C_{39}$ & $2.07\times10^{-6}$ (30) & $5.65\times10^{-7}$ (30) & $2.69\times10^{-6}$ (30) & $3.57\times10^{-7}$ (30) \\
\hline
$C_{43}$ & $2.53\times10^{-6}$ (30) & $8.00\times10^{-7}$ (30) & $2.50\times10^{-6}$ (30) & $3.25\times10^{-7}$ (30) \\
\hline
$C_{48}$ & $3.38\times10^{-6}$ (30) & $7.54\times10^{-7}$ (30) & $4.19\times10^{-6}$ (30) & $3.48\times10^{-7}$ (30) \\
\hline
$C_{50}$ & $3.82\times10^{-6}$ (30) & $8.65\times10^{-7}$ (30) & $2.90\times10^{-6}$ (30) & $4.12\times10^{-7}$ (30) \\
\hline
$C_{57}$ & $2.31\times10^{-6}$ (30) & $8.32\times10^{-7}$ (30) & $2.90\times10^{-6}$ (30) & $3.61\times10^{-7}$ (30) \\
\hline
$C_{60}$ & $2.76\times10^{-6}$ (27) & $1.48\times10^{-6}$ (29) & $2.50\times10^{-6}$ (24) & $8.05\times10^{-8}$ (15) \\
\hline
$C_{62}$ & $3.23\times10^{-6}$ (30) & $8.58\times10^{-7}$ (30) & $3.10\times10^{-6}$ (30) & $4.24\times10^{-7}$ (30) \\
\hline
$C_{65}$ & $1.66\times10^{-6}$ (30) & $6.98\times10^{-7}$ (30) & $2.83\times10^{-6}$ (30) & $4.17\times10^{-7}$ (30) \\
\hline
$C_{67}$ & $3.65\times10^{-6}$ (30) & $6.18\times10^{-7}$ (30) & $1.98\times10^{-6}$ (30) & $3.88\times10^{-7}$ (30) \\
\hline
$C_{68}$ & $2.45\times10^{-6}$ (30) & $8.74\times10^{-7}$ (30) & $2.97\times10^{-6}$ (30) & $4.34\times10^{-7}$ (30) \\
\hline
$C_{72}$ & $2.57\times10^{-6}$ (30) & $9.96\times10^{-7}$ (30) & $3.01\times10^{-6}$ (30) & $3.85\times10^{-7}$ (30) \\
\hline
$C_{75}$ & $3.76\times10^{-6}$ (30) & $7.14\times10^{-7}$ (30) & $2.82\times10^{-6}$ (30) & $3.27\times10^{-7}$ (30) \\
\hline
$C_{80}$ & $4.16\times10^{-6}$ (29) & $8.72\times10^{-7}$ (30) & $2.91\times10^{-6}$ (30) & $3.86\times10^{-7}$ (30) \\
\hline
$C_{82}$ & $2.16\times10^{-6}$ (30) & $6.96\times10^{-7}$ (30) & $3.06\times10^{-6}$ (30) & $3.10\times10^{-7}$ (30) \\
\hline
\end{tabular}
\captionof{table}{Average percentage error in the final energy obtained using qubit-ADAPT-VQE over 100 randomly LU-equivalent states from each entanglement class, evaluated under considered spin models. Values in parentheses correspond to the average number of operators added to ansatz to reach convergence.}
\label{table_100}
\end{widetext}

\subsection{Entangled ground state simulation with VQE}

In the previous sections we use the qubit-ADAPT-VQE algorithm to simulate entangled ground states, a task known to be difficult since generating strong non local correlations is inherently demanding for variational circuits. We observe that, in general, the representative states of all entanglement classes can converge to the solution, with the convergence behavior depending on the initial energy. This  highly not trivial fact occurs thanks to the gradient criterion, which selects the optimal operator for descending in energy. To study the advantage of this gradient criterion in qubit-ADAPT-VQE, we use the representative states as initial conditions to perform a standard VQE with a fixed ansatz and compare its accuracy with that of the qubit-ADAPT-VQE ansatz. We use the EfficientSU2 ansatz \cite{EfficientSU2} from Qiskit library. From numerical simulations we verify that three repetitions of the ansatz are necessary, leading to a set of 32 parameters to optimize. Table \ref{table_VQE} shows the percentage error obtained when we initialize with the representative states of the entanglement classes. We observe a lower performance in both percentage error and number of parameters when compared with qubit-ADAPT-VQE in all models, except for the XXZ case with $\Delta = -0.1$. However, when considering Table \ref{table_100}, where different initial energies are included, the average performance of qubit-ADAPT-VQE remains superior. 

\begin{widetext}
\centering
\begin{tabular}{|c|c|c|c|c|}
\hline
\textbf{Class} & 
\textbf{XY, $\gamma=1$ } & 
\textbf{XY, $\gamma=3$ } & 
\textbf{XXZ, $\Delta = -0.1$} & 
\textbf{XXZ, $\Delta = 3$ } \\
\hline
$C_1$  & $3.12\times10^{0}$  & $3.03\times10^{-1}$  & $3.36\times10^{-2}$ & $2.28\times10^{-2}$  \\
\hline
$C_3$  & $3.08\times10^{-5}$  & $3.15\times10^{-6}$  & $6.50\times10^{-2}$  & $2.20\times10^{-2}$ \\
\hline
$C_9$  & $2.20\times 10^{-1}$  & $1.81\times10^{-1}$ & $3.26\times10^{-2}$ & $4.57\times10^{-2}$ \\
\hline
$C_{11}$ & $2.87\times10^{-4}$  & $3.78\times10^{-2}$ & $1.86\times10^{-1}$  & $6.88\times10^{-1}$ \\
\hline
$C_{15}$ & $1.09\times10^{-6}$ & $3.47\times10^{-5}$ & $1.25\times10^{-2}$ & $2.32\times10^{-2}$  \\
\hline
$C_{19}$ & $7.08\times 10^{-1}$  & $5.27\times10^{-3}$  & $3.58\times10^{-1}$  & $2.77\times10^{-2}$  \\
\hline
$C_{22}$ & $1.64\times10^{-2}$  & $7.83\times10^{-3}$  & $1.69\times10^{-1}$ & $7.30\times10^{-2}$ \\
\hline
$C_{26}$ & $1.08\times10^{-2}$  & $1.43\times10^{1}$  & $3.56\times10^{-2}$  & $4.48\times10^{-2}$  \\
\hline
$C_{28}$ & $1.92\times10^{-2}$  & $1.51\times10^{-1}$  & $1.23\times10^{0}$  & $5.22\times10^{-1}$  \\
\hline
$C_{31}$ & $4.30\times10^{-1}$  & $1.96\times10^{-1}$  & $1.10\times10^{0}$  & $1.65\times10^{-2}$ \\
\hline
$C_{33}$ & $1.88\times10^{-2}$  & $2.57\times10^{-1}$  & $2.03\times10^{-1}$  & $1.30\times10^{-3}$ \\
\hline
$C_{36}$ & $6.31\times10^{-1}$  & $2.42\times10^{-2}$  & $2.63\times10^{-1}$  & $2.75\times10^{-1}$ \\
\hline
$C_{39}$ & $5.13\times10^{-3}$  & $2.42\times10^{-2}$  & $3.29\times10^{-3}$  & $1.05\times10^{-1}$  \\
\hline
$C_{43}$ & $5.52\times10^{-2}$  & $6.49\times10^{-2}$  & $4.37\times10^{-3}$ & $4.79\times10^{-2}$  \\
\hline
$C_{48}$ & $6.16\times10^{-2}$  & $3.74\times10^{-2}$  & $1.18\times10^{-3}$  & $1.55\times10^{-3}$  \\
\hline
$C_{50}$ & $4.29 \times10^{-1}$  & $2.36\times10^{-1}$  & $5.25\times10^{-1}$ & $3.70\times10^{-2}$  \\
\hline
$C_{57}$ & $2.16\times10^{-2}$  & $3.16\times10^{-2}$  & $9.03\times10^{-2}$  & $2.92\times10^{-2}$  \\
\hline
$C_{60}$ & $2.84\times10^{-3}$ & $2.27\times10^{-2}$  & $2.23\times10^{-2}$  & $4.02\times10^{-2}$  \\
\hline
$C_{62}$ & $1.20\times10^{-2}$  & $1.35\times10^{-2}$  & $5.29\times10^{-1}$  & $1.05\times10^{-1}$ \\
\hline
$C_{65}$ & $9.33\times10^{-2}$  & $1.32\times10^{-2}$  & $1.99\times10^{-1}$  & $6.87\times10^{-1}$ \\
\hline
$C_{67}$ & $8.06\times10^{-2}$  & $8.81\times10^{-2}$  & $1.95\times10^{-1}$  & $9.32\times10^{-7}$ \\
\hline
$C_{68}$ & $1.17\times10^{-1}$  & $3.23\times10^{-2}$  & $6.05\times10^{-2}$  & $1.31\times10^{-1}$  \\
\hline
$C_{72}$ & $2.22\times10^{-1}$  & $6.76\times10^{-2}$  & $2.70\times10^{-1}$  & $5.73\times10^{-1}$  \\
\hline
$C_{75}$ & $6.36\times10^{-2}$  & $3.37\times10^{-2}$  & $3.50\times10^{-1}$  & $1.26\times10^{-1}$ \\
\hline
$C_{80}$ & $1.59\times10^{-2}$  & $5.55\times10^{-2}$  & $2.15\times10^{-1}$  & $6.31\times10^{-2}$ \\
\hline
$C_{82}$ & $6.62\times10^{-3}$  & $1.02\times10^{-2}$  & $7.39\times10^{-2}$  & $1.57\times10^{-2}$ \\
\hline
\end{tabular}
\captionof{table}{Percentage error in the final energy obtained with standard VQE for the spin models considered. The representative states were used as initial conditions for the EfficientSU2 ansatz with three repetitions, yielding a total of 32 variational parameters.}
\label{table_VQE}
\end{widetext}

\section{Conclusion} 

We studied the qubit-ADAPT-VQE algorithm under different entanglement conditions of the initial state, considering representative states from the algebraic entanglement classes of four-qubit systems. Two well-known spin models, XY and XXZ, were considered due to the distinctive entanglement features of their ground states. We identified six distinct entanglement classes in the ground states of the spin models, including the separable class, for different values of the anisotropy parameters. This enabled us to select two entanglement classes for each model and evaluate the algorithm. We observe that the choice of the entanglement class as the initial state does not have a direct one-to-one correspondence with the algorithm performance in terms of the final percentage error in energy. Instead, other factors such as the initial energy play a more decisive role, both in how close the system starts to the ground state energy and in the possibility of reaching regions with small cost function gradients. Contrary to what it might be expected, initializing the algorithm with the state belonging to the entanglement class of the ground state does not guarantee better performance than starting from any other class. To clearly assess the influence of the initial energy versus the entanglement class, we generated 100 random instances through LU transformations applied to the representative state of each class, which modify the initial state while preserving its entanglement structure. Averaging over these 100 instances in terms of percentage error yields similar results across all entanglement classes for each model studied. These results reveal the versatility of the algorithm to reach an  approximate the ground state, regardless the quantum correlations existing in the initial state. Despite this, we noted that classes $C_9, C_{26}, C_{31}, C_{33}$, and $C_{60}$ accelerate the convergence procedure, requiring a smaller number of operators in the ansatz. This is important because we consistently observe that, although the energy obtained in the last iterations does not change significantly—leading to a lower percentage error in principle—the additional operators  effectively increase the error due to the growing circuit depth. 

Finally, we repeated the analysis by using standard VQE with the EfficientSU2 ansatz. Across all spin models investigated, its performance was consistently lower than that of qubit-ADAPT-VQE. As in the previous algorithm, initializing with the representative state belonging to the entanglement class of the ground state did not yield a smaller relative error. However, in contrast to qubit-ADAPT-VQE, certain representative states led to percentage errors above one percentage, indicating that EfficientSU2 exhibits a bias with respect to the entanglement class. This suggests that not all classes are equally compatible with this ansatz, whereas qubit-ADAPT-VQE can adapt to all of them, achieving superior performance in entanglement-guided state generation.

\section{Acknowledgement}
DT acknowledge grant Posdoctorado UC PD2024-609. HD, VS and DG acknowledge grant FONDECyT Regular nr 1230586, Chile. HD acknowledge ANID BECAS/MAGÍSTER NACIONAL 22251911


\appendix
\section{Algebraic entanglement classes}
\label{appendix1}

To classify the entanglement of the space $\mathcal{H} = \mathcal{H}^{(2)\otimes 4}$, where $\mathcal{H}^{(2)}$ is the Hilbert space of one qubit, Buniy and Kephart \cite{BuniyJPhysAMathTheor2012} propose a classification through the algebraic properties of the space. Under this construction, the division is made under the equivalence classes by the action of the group structure $G =\text{GL}(2,\mathbb{C})^{\times 4} $. Then an equivalence class is defined as

\begin{equation}
    C(\ket{\psi}) = \{\ket{\phi} \in \mathcal{H}: \ket{\phi} = g \ket{\psi}, g \in G\}
\end{equation}

where $\ket{\psi}\in \mathcal{H}$ is called the representative of its class. In the four-qubit case, there are 83 equivalence classes, which reduce to 27 when taking out qubit permutations.\\

In order to distinguish between each equivalence class, a set of 19 discrete invariants is given. By dividing the space into a bipartition such that $\mathcal{H} = \mathcal{H}_A\otimes \mathcal{H}_B$, a state can be expressed as $\ket{\psi} = \sum_{j,k}a_{jk}\ket{j}_A\otimes \ket{k}_B$. For this, all the possible bipartitions of $\mathcal{H}$ must be considered; therefore, let $\mathcal{H}_A = \otimes _{a \in A}\mathcal{H}_a$ and $A \in P'(I)$, where $P'(I)$ represents the power set of $I$  ($I$ represents the set of subsystems in the four-qubits case $I = \{1,2,3,4\}$), but without the empty set and $I$ itself so the bipartition $\mathcal{H} = \mathcal{H}_I\otimes\mathcal{H}_{\varnothing}$ is avoided. Then, discrete invariants are constructed with respect to the two following linear maps 
\begin{equation}
    \begin{gathered}
        f_{A,\ket{\psi}}: \mathcal{H}_A\to \mathcal{H}_B,\qquad f_{A,\ket\psi} (\ket\omega)= (\bra{\omega}\otimes \mathbb{I}_B)\ket{\psi}\\ = \sum_{j,k}a_{jk}\braket{\omega}{j}_A\ket{k}_B\\
        \tilde{f}_{A,\ket{\psi}}: \mathcal{H}\to \mathcal{H}_B\otimes\mathcal{H}_B,\qquad \tilde{f}_{A,\ket\psi} (\ket\omega)=f_{A,\ket\psi} \otimes \mathbb{I}_B
\end{gathered}
\end{equation}

The most important subsets when defining a linear map are the image and the kernel, but only one is needed to know, as they are the orthogonal complements of each other. Then, by considering the dimension of the intersection of the kernels over different bipartition configurations, that is: 

\begin{equation}
\begin{split}
    n_Q(\ket{\psi}) = dim\left(\cap_{A\in Q}\ker f_{A,\ket{\psi}}\right), \quad Q \in P(P'(I))\\
    \tilde{n}_Q(\ket{\psi}) = dim\left(\cap_{A\in Q}\ker \tilde{f}_{A,\ket{\psi}}\right), \quad Q \in P(P'(I))
\end{split}
\end{equation}

the following set is defined:

\begin{equation}
 \tilde{N}'' (\ket{\psi})=   \{m_Q(\ket{\psi})\}_{Q\in R},
 \end{equation}
 where
 \begin{equation}
 m_Q(\ket{\psi}) = \begin{cases}
        n_Q(\ket{\psi}), \quad \cup _{J \in Q}J \subsetneq I \\\tilde n_Q(\ket{\psi}), \quad  \cup _{J \in Q}J = I
    \end{cases}, \quad R \subseteq P ( P'(I))
\end{equation}
$\tilde{N}'' (\ket{\psi})$ corresponds to the set of discrete invariants given to each class. They are discrete, as the dimension is a natural number. In the four-qubit case, as mentioned before, there are only 19 independent invariants given by 19 different sets $Q_m$ where $m = 1,...,19$. The table with the representative of each class and the set of invariants corresponding to it can be seen in tables VIII, IX and X  of \cite{BuniyJPhysAMathTheor2012}.

\section{Representative states of entanglement classes in qubit-ADAPT-VQE}
\label{appendix2}
Using the representative states listed in Table~\ref{table1} as initial conditions, we executed the qubit-ADAPT-VQE algorithm for the XY and XXZ spin models. The results are presented in Table~\ref{table_l}, where we set $\epsilon = 10^{-2}$. In certain cases the algorithm adds no operators to the ansatz (for example, state $C_1$ in the XXZ model with $\Delta = -0.1$), which occurs because the initial state is already an eigenstate of the Hamiltonian and all operator gradients vanish when the variational parameters start at $\theta_k = 0$. We also observe situations in which the algorithm converges to an excited state instead of the ground state, leading to a larger percentage error. Nevertheless, these cases do not compromise our study, since our main interest lies in the entanglement generation capabilities of the method. By examining the LU equivalent states we further confirm that this behavior is not related to the entanglement class, but rather to the initial energy and the presence of small gradients in the cost function.

\begin{widetext}
\centering
\begin{tabular}{|c|c|c|c|c|}
\hline
\textbf{Class} & 
\textbf{XY, $\gamma=1$ (N° Ops)} & 
\textbf{XY, $\gamma=3$ (N° Ops)} & 
\textbf{XXZ, $\Delta = -0.1$ (N° Ops)} & 
\textbf{XXZ, $\Delta = 3$ (N° Ops)} \\
\hline
$C_1$ & $8.17\times10^{-6}$ (7) & $8.34\times10^{-6}$ (7) & $1.60\times10^{1}$ (0) & $2.12\times10^{0}$ (0) \\
\hline
$C_3$ & $8.43\times10^{-6}$ (7) & $1.29\times10^{-6}$ (8) & $1.30\times10^{0}$ (4) & $1.42\times10^{-5}$ (12) \\
\hline
$C_9$ & $7.51\times10^{-6}$ (6) & $8.97\times10^{-6}$ (7) & $1.20\times10^{0}$ (3) & $4.77\times10^{-6}$ (3) \\
\hline
$C_{11}$ & $7.90\times10^{-6}$ (6) & $1.66\times10^{-6}$ (7) & $1.60\times10^{0}$ (7) & $1.42\times10^{-5}$ (12) \\
\hline
$C_{15}$ & $1.36\times10^{1}$ (6) & $1.46\times10^{-5}$ (11) & $7.49\times10^{-6}$ (6) & $1.61\times10^{-5}$ (13) \\
\hline
$C_{19}$ & $1.18\times10^{-6}$ (3) & $9.05\times10^{-6}$ (8) & $1.50\times10^{0}$ (6) & $1.13\times10^{-5}$ (8) \\
\hline
$C_{22}$ & $1.71\times10^{-4}$ (16) & $1.95\times10^{-5}$ (16) & $1.54\times10^{-5}$ (10) & $1.57\times10^{-5}$ (14) \\
\hline
$C_{26}$ & $8.41\times10^{-6}$ (6) & $8.47\times10^{-6}$ (7) & $3.67\times10^{0}$ (1) & $2.56\times10^{2}$ (1) \\
\hline
$C_{28}$ & $8.04\times10^{-6}$ (7) & $9.23\times10^{-6}$ (7) & $1.60\times10^{0}$ (7) & $1.75\times10^{-5}$ (16) \\
\hline
$C_{31}$ & $2.40\times10^{-5}$ (15) & $1.79\times10^{-5}$ (16) & $1.70\times10^{0}$ (8) & $1.18\times10^{-6}$ (5) \\
\hline
$C_{33}$ & $1.02\times10^{-5}$ (6) & $9.67\times10^{-6}$ (7) & $1.20\times10^{0}$ (3) & $5.87\times10^{-7}$ (3) \\
\hline
$C_{36}$ & $2.10\times10^{-5}$ (16) & $2.37\times10^{-6}$ (16) & $1.99\times10^{-5}$ (15) & $1.69\times10^{-5}$ (15) \\
\hline
$C_{39}$ & $1.14\times10^{-5}$ (7) & $9.57\times10^{-7}$ (7) & $1.60\times10^{0}$ (7) & $2.57\times10^{-6}$ (17) \\
\hline
$C_{43}$ & $1.99\times10^{-5}$ (15) & $1.73\times10^{-5}$ (16) & $2.22\times10^{-5}$ (15) & $2.30\times10^{-5}$ (21) \\
\hline
$C_{48}$ & $1.40\times10^{-4}$ (12) & $1.31\times10^{-5}$ (11) & $1.61\times10^{-4}$ (15) & $1.09\times10^{-5}$ (8) \\
\hline
$C_{50}$ & $1.66\times10^{1}$ (9) & $2.00\times10^{-5}$ (17) & $5.16\times10^{-6}$ (4) & $2.34\times10^{-5}$ (21) \\
\hline
$C_{57}$ & $1.09\times10^{-5}$ (6) & $1.26\times10^{-5}$ (8) & $1.50\times10^{0}$ (6) & $1.21\times10^{-5}$ (9) \\
\hline
$C_{60}$ & $1.05\times10^{-5}$ (7) & $8.64\times10^{-6}$ (7) & $7.67\times10^{0}$ (5) & $4.74\times10^{-6}$ (3) \\
\hline
$C_{62}$ & $2.15\times10^{-5}$ (15) & $2.30\times10^{-5}$ (17) & $2.20\times10^{-5}$ (16) & $3.13\times10^{-5}$ (20) \\
\hline
$C_{65}$ & $1.68\times10^{-4}$ (15) & $2.05\times10^{-5}$ (17) & $1.76\times10^{-4}$ (16) & $2.18\times10^{-5}$ (20) \\
\hline
$C_{67}$ & $1.78\times10^{-4}$ (15) & $1.77\times10^{-5}$ (16) & $1.72\times10^{-4}$ (15) & $2.39\times10^{-5}$ (22) \\
\hline
$C_{68}$ & $1.70\times10^{-4}$ (15) & $1.81\times10^{-5}$ (16) & $1.65\times10^{-4}$ (15) & $2.15\times10^{-5}$ (20) \\
\hline
$C_{72}$ & $1.66\times10^{1}$ (9) & $2.01\times10^{-5}$ (17) & $1.25\times10^{-5}$ (10) & $1.62\times10^{-5}$ (15) \\
\hline
$C_{75}$ & $1.77\times10^{-4}$ (16) & $2.31\times10^{-5}$ (17) & $1.84\times10^{-4}$ (16) & $1.53\times10^{-5}$ (14) \\
\hline
$C_{80}$ & $1.68\times10^{-4}$ (15) & $1.86\times10^{-5}$ (16) & $1.76\times10^{-5}$ (15) & $1.98\times10^{-5}$ (18) \\
\hline
$C_{82}$ & $1.59\times10^{-4}$ (14) & $1.89\times10^{-5}$ (16) & $2.37\times10^{-5}$ (16) & $1.88\times10^{-5}$ (17) \\
\hline
\end{tabular}
\captionof{table}{Percentage error of final energy obtained with qubit-ADAPT-VQE over representative state of entanglement classes under different spin models. The number in parentheses indicates the number of operators added to the ansatz.}
\label{table_l}
\end{widetext}

\end{document}